\documentclass[12pt]{article}
\usepackage{epsfig,amsfonts,amssymb}
\usepackage{hyperref}
\usepackage{cite}
\input epsf.sty
\usepackage{cite}

\def\ZZZ{{\hbox{ Z\kern-1.6mm Z}}}
\def\RRR{{\hbox{ R\kern-2.4mm R}}}
\def\CCC{{\hbox{ C\kern-2.0mm C}}}
\def\zzz{{\hbox{z\kern-1mm z}}}

\newcommand{\vt}{\vartheta}

\newcommand{\qeq}{{\hbox{=\kern-2.3mm ? \kern.5mm }}}
\renewcommand{\qeq}{=}

\newcommand{\eps}{\epsilon}

\newcommand{\OO}{{\cal O}}

\newcommand{\wt}{\widetilde}

\newcommand{\be}{\begin{equation}}
\newcommand{\ee}{\end{equation}}
\newcommand{\ben}{\begin{eqnarray}\displaystyle}
\newcommand{\een}{\end{eqnarray}}

\newcommand{\refb}[1]{(\ref{#1})}
\newcommand{\p}{\partial}
\newcommand{\sectiono}[1]{\section{#1}\setcounter{equation}{0}}

\def\one{{\hbox{ 1\kern-.8mm l}}}
\def\zero{{\hbox{ 0\kern-1.5mm 0}}}

\newcommand{\bea}[1]{\begin{eqnarray}\label{#1} }
\newcommand{\eea}{\end{eqnarray}}

\newcommand{\eqref}{\refb}

\usepackage{bm}
\usepackage[table]{xcolor}

\def\figeps{

\def\JPicScale{0.8}
\ifx\JPicScale\undefined\def\JPicScale{1}\fi
\unitlength \JPicScale mm
\begin{picture}(150,90)(0,0)
\linethickness{0.3mm}
\multiput(70,0)(0,1.98){46}{\line(0,1){0.99}}
\linethickness{0.35mm}
\multiput(10,50.5)(1.98,0){66}{\line(1,0){0.99}}
\linethickness{0.3mm}
\put(40,0){\line(0,1){45}}
\linethickness{0.3mm}
\multiput(40,45)(0.66,0.12){167}{\line(1,0){0.66}}
\linethickness{0.3mm}
\put(150,65){\line(0,1){25}}
\put(50,55){\makebox(0,0)[cc]{$\bf Q_2$}}

\put(50,50.5){\makebox(0,0)[cc]{$\times$}}

\put(90,45){\makebox(0,0)[cc]{$\bf Q_1$}}

\put(90,50.5){\makebox(0,0)[cc]{$\times$}}

\put(100,65){\makebox(0,0)[cc]{$\bf Q_4$}}

\put(100,59.5){\makebox(0,0)[cc]{$\times$}}

\put(120,65){\makebox(0,0)[cc]{$\bf p^0$}}

\put(120,59.5){\makebox(0,0)[cc]{$\times$}}

\put(140,55){\makebox(0,0)[cc]{$\bf Q_3$}}

\put(140,59.5){\makebox(0,0)[cc]{$\times$}}

\end{picture}

}

\def\figqft{

\def\JPicScale{0.8}
\ifx\JPicScale\undefined\def\JPicScale{1}\fi
\unitlength \JPicScale mm
\begin{picture}(135,80)(0,0)
\linethickness{0.3mm}
\put(100.35,48.61){\line(0,1){0.5}}
\multiput(100.1,52.12)(0.07,-0.5){1}{\line(0,-1){0.5}}
\multiput(100.02,52.62)(0.08,-0.5){1}{\line(0,-1){0.5}}
\multiput(99.92,53.11)(0.1,-0.49){1}{\line(0,-1){0.49}}
\multiput(99.82,53.61)(0.11,-0.49){1}{\line(0,-1){0.49}}
\multiput(99.7,54.09)(0.12,-0.49){1}{\line(0,-1){0.49}}
\multiput(98.93,56.49)(0.18,-0.47){1}{\line(0,-1){0.47}}
\multiput(98.75,56.96)(0.09,-0.23){2}{\line(0,-1){0.23}}
\multiput(98.55,57.42)(0.1,-0.23){2}{\line(0,-1){0.23}}
\multiput(98.34,57.88)(0.1,-0.23){2}{\line(0,-1){0.23}}
\multiput(97.13,60.09)(0.13,-0.21){2}{\line(0,-1){0.21}}
\multiput(96.86,60.51)(0.14,-0.21){2}{\line(0,-1){0.21}}
\multiput(96.58,60.93)(0.14,-0.21){2}{\line(0,-1){0.21}}
\multiput(96.29,61.34)(0.15,-0.21){2}{\line(0,-1){0.21}}
\multiput(94.69,63.28)(0.11,-0.12){3}{\line(0,-1){0.12}}
\multiput(94.34,63.64)(0.12,-0.12){3}{\line(0,-1){0.12}}
\multiput(93.98,64)(0.12,-0.12){3}{\line(0,-1){0.12}}
\multiput(93.62,64.34)(0.12,-0.12){3}{\line(1,0){0.12}}
\multiput(91.68,65.95)(0.13,-0.1){3}{\line(1,0){0.13}}
\multiput(91.27,66.24)(0.21,-0.15){2}{\line(1,0){0.21}}
\multiput(90.85,66.52)(0.21,-0.14){2}{\line(1,0){0.21}}
\multiput(90.43,66.79)(0.21,-0.14){2}{\line(1,0){0.21}}
\multiput(88.22,68)(0.23,-0.11){2}{\line(1,0){0.23}}
\multiput(87.76,68.21)(0.23,-0.1){2}{\line(1,0){0.23}}
\multiput(87.3,68.4)(0.23,-0.1){2}{\line(1,0){0.23}}
\multiput(86.83,68.59)(0.23,-0.09){2}{\line(1,0){0.23}}
\multiput(84.44,69.36)(0.49,-0.13){1}{\line(1,0){0.49}}
\multiput(83.95,69.48)(0.49,-0.12){1}{\line(1,0){0.49}}
\multiput(83.45,69.58)(0.49,-0.11){1}{\line(1,0){0.49}}
\multiput(82.96,69.68)(0.49,-0.1){1}{\line(1,0){0.49}}
\multiput(80.46,69.98)(0.5,-0.04){1}{\line(1,0){0.5}}
\multiput(79.96,70)(0.5,-0.02){1}{\line(1,0){0.5}}
\multiput(79.46,70.01)(0.5,-0.01){1}{\line(1,0){0.5}}
\put(78.95,70.01){\line(1,0){0.5}}
\multiput(76.44,69.83)(0.5,0.06){1}{\line(1,0){0.5}}
\multiput(75.95,69.76)(0.5,0.07){1}{\line(1,0){0.5}}
\multiput(75.45,69.68)(0.5,0.08){1}{\line(1,0){0.5}}
\multiput(74.95,69.58)(0.49,0.1){1}{\line(1,0){0.49}}
\multiput(72.52,68.93)(0.48,0.15){1}{\line(1,0){0.48}}
\multiput(72.05,68.77)(0.48,0.16){1}{\line(1,0){0.48}}
\multiput(71.58,68.59)(0.47,0.18){1}{\line(1,0){0.47}}
\multiput(71.11,68.4)(0.23,0.09){2}{\line(1,0){0.23}}
\multiput(68.85,67.31)(0.22,0.12){2}{\line(1,0){0.22}}
\multiput(68.41,67.05)(0.22,0.13){2}{\line(1,0){0.22}}
\multiput(67.98,66.79)(0.21,0.13){2}{\line(1,0){0.21}}
\multiput(67.56,66.52)(0.21,0.14){2}{\line(1,0){0.21}}
\multiput(65.54,65.01)(0.13,0.11){3}{\line(1,0){0.13}}
\multiput(65.16,64.68)(0.13,0.11){3}{\line(1,0){0.13}}
\multiput(64.79,64.34)(0.12,0.11){3}{\line(1,0){0.12}}
\multiput(64.43,64)(0.12,0.12){3}{\line(1,0){0.12}}
\multiput(62.74,62.14)(0.11,0.13){3}{\line(0,1){0.13}}
\multiput(62.42,61.74)(0.1,0.13){3}{\line(0,1){0.13}}
\multiput(62.12,61.34)(0.1,0.13){3}{\line(0,1){0.13}}
\multiput(61.83,60.93)(0.15,0.21){2}{\line(0,1){0.21}}
\multiput(60.52,58.78)(0.12,0.22){2}{\line(0,1){0.22}}
\multiput(60.29,58.33)(0.12,0.22){2}{\line(0,1){0.22}}
\multiput(60.07,57.88)(0.11,0.23){2}{\line(0,1){0.23}}
\multiput(59.86,57.42)(0.1,0.23){2}{\line(0,1){0.23}}
\multiput(58.98,55.06)(0.15,0.48){1}{\line(0,1){0.48}}
\multiput(58.84,54.58)(0.14,0.48){1}{\line(0,1){0.48}}
\multiput(58.71,54.09)(0.13,0.49){1}{\line(0,1){0.49}}
\multiput(58.59,53.61)(0.12,0.49){1}{\line(0,1){0.49}}
\multiput(58.17,51.12)(0.06,0.5){1}{\line(0,1){0.5}}
\multiput(58.13,50.62)(0.05,0.5){1}{\line(0,1){0.5}}
\multiput(58.09,50.12)(0.04,0.5){1}{\line(0,1){0.5}}
\multiput(58.07,49.62)(0.02,0.5){1}{\line(0,1){0.5}}
\multiput(58.05,49.12)(0.01,0.5){1}{\line(0,1){0.5}}
\put(58.05,48.61){\line(0,1){0.5}}
\multiput(58.05,48.61)(0.01,-0.5){1}{\line(0,-1){0.5}}
\multiput(58.07,48.11)(0.02,-0.5){1}{\line(0,-1){0.5}}
\multiput(58.09,47.61)(0.04,-0.5){1}{\line(0,-1){0.5}}
\multiput(58.13,47.1)(0.05,-0.5){1}{\line(0,-1){0.5}}
\multiput(58.17,46.6)(0.06,-0.5){1}{\line(0,-1){0.5}}
\multiput(58.23,46.1)(0.07,-0.5){1}{\line(0,-1){0.5}}
\multiput(58.31,45.6)(0.08,-0.5){1}{\line(0,-1){0.5}}
\multiput(58.39,45.11)(0.1,-0.49){1}{\line(0,-1){0.49}}
\multiput(58.48,44.61)(0.11,-0.49){1}{\line(0,-1){0.49}}
\multiput(58.59,44.12)(0.12,-0.49){1}{\line(0,-1){0.49}}
\multiput(58.71,43.63)(0.13,-0.49){1}{\line(0,-1){0.49}}
\multiput(58.84,43.15)(0.14,-0.48){1}{\line(0,-1){0.48}}
\multiput(58.98,42.66)(0.15,-0.48){1}{\line(0,-1){0.48}}
\multiput(59.14,42.18)(0.16,-0.48){1}{\line(0,-1){0.48}}
\multiput(59.3,41.71)(0.18,-0.47){1}{\line(0,-1){0.47}}
\multiput(59.48,41.24)(0.09,-0.23){2}{\line(0,-1){0.23}}
\multiput(59.66,40.77)(0.1,-0.23){2}{\line(0,-1){0.23}}
\multiput(59.86,40.31)(0.1,-0.23){2}{\line(0,-1){0.23}}
\multiput(60.07,39.85)(0.11,-0.23){2}{\line(0,-1){0.23}}
\multiput(60.29,39.4)(0.12,-0.22){2}{\line(0,-1){0.22}}
\multiput(60.52,38.95)(0.12,-0.22){2}{\line(0,-1){0.22}}
\multiput(60.76,38.51)(0.13,-0.22){2}{\line(0,-1){0.22}}
\multiput(61.01,38.07)(0.13,-0.21){2}{\line(0,-1){0.21}}
\multiput(61.28,37.64)(0.14,-0.21){2}{\line(0,-1){0.21}}
\multiput(61.55,37.22)(0.14,-0.21){2}{\line(0,-1){0.21}}
\multiput(61.83,36.8)(0.15,-0.21){2}{\line(0,-1){0.21}}
\multiput(62.12,36.39)(0.1,-0.13){3}{\line(0,-1){0.13}}
\multiput(62.42,35.99)(0.1,-0.13){3}{\line(0,-1){0.13}}
\multiput(62.74,35.59)(0.11,-0.13){3}{\line(0,-1){0.13}}
\multiput(63.06,35.2)(0.11,-0.13){3}{\line(0,-1){0.13}}
\multiput(63.39,34.82)(0.11,-0.12){3}{\line(0,-1){0.12}}
\multiput(63.72,34.45)(0.12,-0.12){3}{\line(0,-1){0.12}}
\multiput(64.07,34.09)(0.12,-0.12){3}{\line(0,-1){0.12}}
\multiput(64.43,33.73)(0.12,-0.12){3}{\line(1,0){0.12}}
\multiput(64.79,33.38)(0.12,-0.11){3}{\line(1,0){0.12}}
\multiput(65.16,33.04)(0.13,-0.11){3}{\line(1,0){0.13}}
\multiput(65.54,32.71)(0.13,-0.11){3}{\line(1,0){0.13}}
\multiput(65.93,32.39)(0.13,-0.1){3}{\line(1,0){0.13}}
\multiput(66.33,32.08)(0.13,-0.1){3}{\line(1,0){0.13}}
\multiput(66.73,31.78)(0.21,-0.15){2}{\line(1,0){0.21}}
\multiput(67.14,31.49)(0.21,-0.14){2}{\line(1,0){0.21}}
\multiput(67.56,31.21)(0.21,-0.14){2}{\line(1,0){0.21}}
\multiput(67.98,30.94)(0.21,-0.13){2}{\line(1,0){0.21}}
\multiput(68.41,30.67)(0.22,-0.13){2}{\line(1,0){0.22}}
\multiput(68.85,30.42)(0.22,-0.12){2}{\line(1,0){0.22}}
\multiput(69.29,30.18)(0.22,-0.12){2}{\line(1,0){0.22}}
\multiput(69.74,29.95)(0.23,-0.11){2}{\line(1,0){0.23}}
\multiput(70.19,29.73)(0.23,-0.1){2}{\line(1,0){0.23}}
\multiput(70.65,29.52)(0.23,-0.1){2}{\line(1,0){0.23}}
\multiput(71.11,29.32)(0.23,-0.09){2}{\line(1,0){0.23}}
\multiput(71.58,29.14)(0.47,-0.18){1}{\line(1,0){0.47}}
\multiput(72.05,28.96)(0.48,-0.16){1}{\line(1,0){0.48}}
\multiput(72.52,28.79)(0.48,-0.15){1}{\line(1,0){0.48}}
\multiput(73,28.64)(0.48,-0.14){1}{\line(1,0){0.48}}
\multiput(73.49,28.5)(0.49,-0.13){1}{\line(1,0){0.49}}
\multiput(73.97,28.37)(0.49,-0.12){1}{\line(1,0){0.49}}
\multiput(74.46,28.25)(0.49,-0.11){1}{\line(1,0){0.49}}
\multiput(74.95,28.14)(0.49,-0.1){1}{\line(1,0){0.49}}
\multiput(75.45,28.05)(0.5,-0.08){1}{\line(1,0){0.5}}
\multiput(75.95,27.96)(0.5,-0.07){1}{\line(1,0){0.5}}
\multiput(76.44,27.89)(0.5,-0.06){1}{\line(1,0){0.5}}
\multiput(76.94,27.83)(0.5,-0.05){1}{\line(1,0){0.5}}
\multiput(77.44,27.79)(0.5,-0.04){1}{\line(1,0){0.5}}
\multiput(77.95,27.75)(0.5,-0.02){1}{\line(1,0){0.5}}
\multiput(78.45,27.73)(0.5,-0.01){1}{\line(1,0){0.5}}
\put(78.95,27.71){\line(1,0){0.5}}
\multiput(79.46,27.71)(0.5,0.01){1}{\line(1,0){0.5}}
\multiput(79.96,27.73)(0.5,0.02){1}{\line(1,0){0.5}}
\multiput(80.46,27.75)(0.5,0.04){1}{\line(1,0){0.5}}
\multiput(80.96,27.79)(0.5,0.05){1}{\line(1,0){0.5}}
\multiput(81.47,27.83)(0.5,0.06){1}{\line(1,0){0.5}}
\multiput(81.97,27.89)(0.5,0.07){1}{\line(1,0){0.5}}
\multiput(82.46,27.96)(0.5,0.08){1}{\line(1,0){0.5}}
\multiput(82.96,28.05)(0.49,0.1){1}{\line(1,0){0.49}}
\multiput(83.45,28.14)(0.49,0.11){1}{\line(1,0){0.49}}
\multiput(83.95,28.25)(0.49,0.12){1}{\line(1,0){0.49}}
\multiput(84.44,28.37)(0.49,0.13){1}{\line(1,0){0.49}}
\multiput(84.92,28.5)(0.48,0.14){1}{\line(1,0){0.48}}
\multiput(85.4,28.64)(0.48,0.15){1}{\line(1,0){0.48}}
\multiput(85.88,28.79)(0.48,0.16){1}{\line(1,0){0.48}}
\multiput(86.36,28.96)(0.47,0.18){1}{\line(1,0){0.47}}
\multiput(86.83,29.14)(0.23,0.09){2}{\line(1,0){0.23}}
\multiput(87.3,29.32)(0.23,0.1){2}{\line(1,0){0.23}}
\multiput(87.76,29.52)(0.23,0.1){2}{\line(1,0){0.23}}
\multiput(88.22,29.73)(0.23,0.11){2}{\line(1,0){0.23}}
\multiput(88.67,29.95)(0.22,0.12){2}{\line(1,0){0.22}}
\multiput(89.12,30.18)(0.22,0.12){2}{\line(1,0){0.22}}
\multiput(89.56,30.42)(0.22,0.13){2}{\line(1,0){0.22}}
\multiput(90,30.67)(0.21,0.13){2}{\line(1,0){0.21}}
\multiput(90.43,30.94)(0.21,0.14){2}{\line(1,0){0.21}}
\multiput(90.85,31.21)(0.21,0.14){2}{\line(1,0){0.21}}
\multiput(91.27,31.49)(0.21,0.15){2}{\line(1,0){0.21}}
\multiput(91.68,31.78)(0.13,0.1){3}{\line(1,0){0.13}}
\multiput(92.08,32.08)(0.13,0.1){3}{\line(1,0){0.13}}
\multiput(92.48,32.39)(0.13,0.11){3}{\line(1,0){0.13}}
\multiput(92.86,32.71)(0.13,0.11){3}{\line(1,0){0.13}}
\multiput(93.24,33.04)(0.12,0.11){3}{\line(1,0){0.12}}
\multiput(93.62,33.38)(0.12,0.12){3}{\line(1,0){0.12}}
\multiput(93.98,33.73)(0.12,0.12){3}{\line(1,0){0.12}}
\multiput(94.34,34.09)(0.12,0.12){3}{\line(0,1){0.12}}
\multiput(94.69,34.45)(0.11,0.12){3}{\line(0,1){0.12}}
\multiput(95.02,34.82)(0.11,0.13){3}{\line(0,1){0.13}}
\multiput(95.35,35.2)(0.11,0.13){3}{\line(0,1){0.13}}
\multiput(95.67,35.59)(0.1,0.13){3}{\line(0,1){0.13}}
\multiput(95.99,35.99)(0.1,0.13){3}{\line(0,1){0.13}}
\multiput(96.29,36.39)(0.15,0.21){2}{\line(0,1){0.21}}
\multiput(96.58,36.8)(0.14,0.21){2}{\line(0,1){0.21}}
\multiput(96.86,37.22)(0.14,0.21){2}{\line(0,1){0.21}}
\multiput(97.13,37.64)(0.13,0.21){2}{\line(0,1){0.21}}
\multiput(97.39,38.07)(0.13,0.22){2}{\line(0,1){0.22}}
\multiput(97.65,38.51)(0.12,0.22){2}{\line(0,1){0.22}}
\multiput(97.89,38.95)(0.12,0.22){2}{\line(0,1){0.22}}
\multiput(98.12,39.4)(0.11,0.23){2}{\line(0,1){0.23}}
\multiput(98.34,39.85)(0.1,0.23){2}{\line(0,1){0.23}}
\multiput(98.55,40.31)(0.1,0.23){2}{\line(0,1){0.23}}
\multiput(98.75,40.77)(0.09,0.23){2}{\line(0,1){0.23}}
\multiput(98.93,41.24)(0.18,0.47){1}{\line(0,1){0.47}}
\multiput(99.11,41.71)(0.16,0.48){1}{\line(0,1){0.48}}
\multiput(99.27,42.18)(0.15,0.48){1}{\line(0,1){0.48}}
\multiput(99.43,42.66)(0.14,0.48){1}{\line(0,1){0.48}}
\multiput(99.57,43.15)(0.13,0.49){1}{\line(0,1){0.49}}
\multiput(99.7,43.63)(0.12,0.49){1}{\line(0,1){0.49}}
\multiput(99.82,44.12)(0.11,0.49){1}{\line(0,1){0.49}}
\multiput(99.92,44.61)(0.1,0.49){1}{\line(0,1){0.49}}
\multiput(100.02,45.11)(0.08,0.5){1}{\line(0,1){0.5}}
\multiput(100.1,45.6)(0.07,0.5){1}{\line(0,1){0.5}}
\multiput(100.17,46.1)(0.06,0.5){1}{\line(0,1){0.5}}
\multiput(100.23,46.6)(0.05,0.5){1}{\line(0,1){0.5}}
\multiput(100.28,47.1)(0.04,0.5){1}{\line(0,1){0.5}}
\multiput(100.32,47.61)(0.02,0.5){1}{\line(0,1){0.5}}
\multiput(100.34,48.11)(0.01,0.5){1}{\line(0,1){0.5}}

\linethickness{0.9mm}
\put(30,50){\line(1,0){28}}
\linethickness{0.9mm}
\put(100,50){\line(1,0){28}}
\put(40,55){\makebox(0,0)[cc]{$p$}}

\put(115,55){\makebox(0,0)[cc]{$p$}}

\put(78,77){\makebox(0,0)[cc]{$k$}}

\put(79,21){\makebox(0,0)[cc]{$p-k$}}

\end{picture}

}

\def\figtwo{

\def\JPicScale{0.5}
\ifx\JPicScale\undefined\def\JPicScale{1}\fi
\unitlength \JPicScale mm
\begin{picture}(155,90)(0,0)
\linethickness{0.3mm}
\put(80,0){\line(0,1){40}}
\linethickness{0.3mm}
\put(80,40){\line(1,0){40}}
\linethickness{0.3mm}
\put(120,40){\line(0,1){20}}
\linethickness{0.3mm}
\put(80,60){\line(1,0){40}}
\linethickness{0.3mm}
\put(80,60){\line(0,1){30}}
\put(140,50){\makebox(0,0)[cc]{x}}

\put(155,50){\makebox(0,0)[cc]{x}}

\put(20,50){\makebox(0,0)[cc]{x}}

\put(100,50){\makebox(0,0)[cc]{x}}

\put(15,45){\makebox(0,0)[cc]{$Q_2$}}

\put(140,45){\makebox(0,0)[cc]{$Q_1$}}

\put(100,45){\makebox(0,0)[cc]{$Q_4$}}

\put(155,45){\makebox(0,0)[cc]{$Q_3$}}

\put(80,-10){\makebox(0,0)[cc]{(a)}}

\end{picture}

}

\def\figfour{

\def\JPicScale{0.5}
\ifx\JPicScale\undefined\def\JPicScale{1}\fi
\unitlength \JPicScale mm
\begin{picture}(150,90)(0,0)
\put(30,50){\makebox(0,0)[cc]{x}}

\put(90,50){\makebox(0,0)[cc]{x}}

\put(110,50){\makebox(0,0)[cc]{x}}

\put(150,50){\makebox(0,0)[cc]{x}}

\put(25,45){\makebox(0,0)[cc]{$Q_2$}}

\put(85,45){\makebox(0,0)[cc]{$Q_1$}}

\put(115,45){\makebox(0,0)[cc]{$Q_4$}}

\put(150,45){\makebox(0,0)[cc]{$Q_3$}}

\put(80,-10){\makebox(0,0)[cc]{(b)}}

\linethickness{0.3mm}
\qbezier(60,90)(60,84.8)(60,81.19)
\qbezier(60,81.19)(60,77.58)(60,75)
\qbezier(60,75)(60,72.41)(60,70)
\qbezier(60,70)(60,67.59)(60,65)
\qbezier(60,65)(59.98,62.4)(61.19,60.59)
\qbezier(61.19,60.59)(62.39,58.79)(65,57.5)
\qbezier(65,57.5)(67.59,56.16)(70.59,58.56)
\qbezier(70.59,58.56)(73.6,60.97)(77.5,67.5)
\qbezier(77.5,67.5)(81.38,74.02)(85.59,77.62)
\qbezier(85.59,77.62)(89.8,81.23)(95,82.5)
\qbezier(95,82.5)(100.2,83.83)(104.41,82.62)
\qbezier(104.41,82.62)(108.62,81.42)(112.5,77.5)
\qbezier(112.5,77.5)(116.41,73.66)(118.22,66.44)
\qbezier(118.22,66.44)(120.02,59.22)(120,47.5)
\qbezier(120,47.5)(120.04,35.69)(117.03,35.69)
\qbezier(117.03,35.69)(114.02,35.69)(107.5,47.5)
\qbezier(107.5,47.5)(101.03,59.29)(93.81,61.09)

\qbezier(93.81,61.09)(86.59,62.9)(77.5,55)
\qbezier(77.5,55)(68.37,47.22)(64.16,40)
\qbezier(64.16,40)(59.95,32.78)(60,25)
\qbezier(60,25)(60,17.19)(60,12.38)
\qbezier(60,12.38)(60,7.56)(60,5)
\qbezier(60,5)(60,2.41)(60,0)
\qbezier(60,0)(60,-2.41)(60,-5)
\end{picture}

}

\begin{document}

\vskip 12pt

\baselineskip 24pt

\begin{center}
{\Large \bf  One Loop Mass Renormalization of Unstable Particles in Superstring Theory}

\end{center}


\vspace*{2.0ex}

\baselineskip=18pt

\centerline{\large \rm Ashoke Sen}

\vspace*{4.0ex}

\centerline{\large \it Harish-Chandra Research Institute}
\centerline{\large \it  Chhatnag Road, Jhusi,
Allahabad 211019, India}

\centerline{and}

\centerline{\large \it Homi Bhabha National Institute}
\centerline{\large \it Training School Complex, Anushakti Nagar,
    Mumbai 400085, India}

\vspace*{1.0ex}
\centerline{\small E-mail:  sen@mri.ernet.in}

\vspace*{5.0ex}

\centerline{\bf Abstract} \bigskip

Most of the massive states in superstring theory are expected to  undergo mass
renormalization at one loop order. Typically these corrections should contain
imaginary parts, indicating that the states are unstable against decay into
lighter particles. However in such cases,
direct computation of the renormalized mass using superstring
perturbation theory yields divergent result. 
Previous approaches to this problem involve various analytic
continuation techniques, or
deforming the integral
over the moduli space of the torus with two punctures 
into the complexified moduli
space near the boundary.
In this paper we use insights from string field theory to
describe a different approach that gives manifestly finite result 
for the mass shift satisfying unitarity relations.
The procedure is applicable to all states of (compactified)
type II and heterotic string theories.
We illustrate this by computing the
one loop correction to the 
mass of the first massive state  on
the leading Regge trajectory in SO(32) heterotic string theory.

\vfill \eject

\baselineskip 18pt

\tableofcontents

\section{Introduction and Summary} \label{sintro}

The world-sheet formulation of superstring perturbation theory gives an elegant
expression for scattering amplitudes, expressing the amplitude at any given
order in perturbation theory as a single integral over the moduli space of a Riemann
surface with punctures. This expression is manifestly free
from ultraviolet divergences. However superstring perturbation theory shares all
the usual infrared divergence problems in quantum field theory, but unlike in the
case of quantum field theories, there is no systematic way of dealing with these
divergences within the frame-work of the world-sheet formalism.

Superstring field theory provides a solution to this problem. By construction, the
Feynman rules of superstring field theory reproduce the amplitude given by the
world-sheet description when the latter gives finite result, 
but the existence of the underlying quantum field theory
allows us to deal with the infrared divergence problems when they arise. 

In this paper we shall use the insight from superstring field theory to address a
related problem that arises in the world-sheet description of superstring perturbation
theory. String theory has many massive states in its spectrum, but most of them 
are unstable against decay to lighter states. Therefore one expects that when
quantum corrections to the masses are taken into account, the mass$^2$ of an unstable
particle should
receive correction that contains an imaginary part (and also possibly a real part).
Now while higher loop mass renormalization requires full use of string field theory
-- because one needs to subtract the one particle reducible (1PR) contributions
from the two point function -- 
one would expect that the one loop contribution to the shift in mass$^2$ should be
given by the on-shell two point function on the torus, and hence should be
straightforward to compute using the usual world-sheet formalism. 
However when one tries to repeat this
computation for an unstable state, one finds a divergent 
answer\cite{sundborg,amano,sundborg1}.

Intuitively the reason for this divergence is 
as 
follows\cite{sundborg,amano,sundborg1,9302003,9404128,9410152,berera2,1307.5124}. 
In quantum field theory, 
while computing the mass renormalization of a particle that can decay into two
or more particles, one finds that there are Feynman diagrams for which one
or more internal propagators have negative denominator $(k^2+m^2)$ for
some region of internal loop momentum integration, and there is no way to deform
the integration contours of loop
momenta that can make all denominators have positive real parts
everywhere along the contour. This means that the Schwinger parameter 
representation of this propagator breaks down, -- if we try to replace 
$(k^2+m^2)^{-1}$ by $\int_0^\infty ds \, e^{-s (k^2+m^2)}$ then the integration
over $s$ encounters a divergence from infinity.  On the other hand, integration over
the moduli space of Riemann surfaces directly gives the result in the Schwinger
parameter representation. Therefore the issue shows up as a divergence in the
integration over the moduli space of Riemann surfaces. 

It is also possible to argue that a finite result would necessarily
have led to a contradiction. 
The loop correction to 
mass$^2$ of an unstable particle 
is expected to have an imaginary part, but straightforward world-sheet
computation in string theory gives real results for all amplitudes. Therefore
the only way an imaginary part can arise is if the naive world-sheet description
gives divergent answer. In that case one might hope that by defining 
the amplitude for unphysical external
momenta where the result is finite and then analytically continuing the result to
on-shell external momenta, we may get an imaginary part. Early attempts to
implement this achieved only partial
success\cite{sundborg,amano}.
A systematic method of dealing with this was suggested in 
\cite{9302003,9404128,9410152} (see also \cite{amano2,montag}).
This was achieved by considering 
a four point amplitude with external momenta chosen in appropriate range 
where the integrals are well defined,  then
analytically continuing the result to the physical region where we expect a pole
due to the massive particle of interest, and finally finding the shift in mass$^2$ from
the location of the pole. Alternative approaches to analytic continuation, working
directly with two point function, can be found in \cite{marcus,0008060,0210245}.
The imaginary part of the shift, which is related to the decay rate, is
relatively easier to compute, and various other methods for computing this can
be found in \cite{chiu,miransky,green,turok,marcus,dai,okada,
mitchell,9901092,0008060,0109196,0210245}.\footnote{In 
a quantum field theory the imaginary part
is determined by unitarity relation. On the other hand the real part
is ultraviolet divergent. This has to be removed by a counterterm and hence 
has to be taken as an input parameter of the theory. In string
theory both parts are finite and computable.}

One disadvantage of the analytic continuation procedure is that it has to be done on
a case by case basis, and may not provide a systematic procedure to deal with all
cases. For example not every massive state may appear as an
intermediate state in the four point amplitude of massless external states.
Also at higher mass levels there will be mixing between different states, leading
to a renormalized mass$^2$ matrix with both real and imaginary parts, and it 
may not be easy to extract this matrx from the four point function of massless states.
Finally, lack of a general procedure makes it difficult to prove general properties
like unitarity that relates the imaginary part of the mass shift to the decay rate -- except by
explicit computation in each case.
For these reasons,
it will clearly be useful to develop a systematic procedure for computing 
string theory amplitudes that directly gives a finite result instead of having to define
the amplitudes via analytic continuation. This will be
the analog of the $i\eps$ prescription in quantum field theory, -- instead of defining the
amplitudes as the analytic continuations of Euclidean Green's functions, one can write
down the expression for the Green's functions with Lorentzian external momenta as
integrals over loop momenta, but one needs the 
$i\eps$ prescription for regulating the poles of the
propagator.
Proposals for generalizing this to string theory
was given in \cite{berera2,1307.5124}.
These approaches involve
deforming the integration over the moduli space
of Riemann surfaces -- that appear in the expression for the loop amplitudes --
into the complexified moduli space. 
In terms of the Schwinger parameter representation of the
propagators, this corresponds to taking the upper limit of $s$ integration to be
$i\infty$ instead of $\infty$, and at the same time supplying a small
damping factor that represent the effect of replacing $m^2$ by $m^2-i\eps$
as in a conventional quantum field theory.

In this paper we suggest a different approach to 
this problem by directly drawing insight from string field
theory. In any quantum field theory, writing down the expression for a loop amplitude
is quite straightforward if the Feynman rules are known, 
but typically it suffers from ultraviolet
divergence. In string field theory there are no ultraviolet divergences since the
vertices fall off exponentially for large space-like external momenta. 
However in the conventional formulation of string field theory,
the vertices grow exponentially for large time-like momenta.  Due to this
property, while computing Feynman amplitudes by integrating over internal
momenta, we cannot take the integral over internal energies along the real
axis -- the ends of the integration contour have to be tied to 
$\pm i\infty$\cite{1604.01783}. However in the interior of the complex plane the
contour has to be deformed appropriately away from the imaginary axis following the 
algorithm described in \cite{1604.01783}. With this
prescription we get
finite results for all loop corrections except where there are physical infrared divergences 
involving one or more divergent propagators --
{\it e.g.} mass renormalization diagrams if we fail to take into account the shift of mass
due to quantum corrections, or tadpole divergences if the original perturbative vacuum
is destabilized by quantum corrections. In the absence of such divergences, we should get 
finite results. This includes the one loop two point function that is needed for computing
the renormalized mass --  both its real
and the imaginary parts.

One could wonder how the results in string field theory are related to those of other
approaches -- {\it e.g.} analytic continuation. To this end we note that the string field
theory amplitudes, constructed using the procedure mentioned above, are automatically
analytic functions of external momenta. Therefore by the uniqueness of analytic
continuation, string field theory results must agree with those computed using analytic
continuation. However what string field theory achieves is that it expresses the result
as a (contour) integral over momenta that is manifestly finite without any need for
analytic continuation. Therefore this automatically gives the analytically continued 
result that we would have gotten from the usual world-sheet approach. Another bonus
of this approach is that the amplitudes defined this way automatically satisfies
the Cutkosky cutting rules\cite{1604.01783}. 
While for general amplitudes one still needs few
more steps to prove unitarity from the cutting rules by showing that the contribution
to the cut diagrams from unphysical intermediate states cancel, for diagrams involving
one loop mass renormalization this can be shown explicitly. Therefore the imaginary
parts of the mass shifts computed using this approach are automatically consistent
with unitarity.\footnote{Since the approach of \cite{berera2} was motivated from light-cone
string field theory\cite{mandelstam1,mandelstam2}, 
one could ask if we can directly work with the light-cone string
field theory and impose the $i\eps$ prescription there. This would make the 
proof of unitarity more straightforward.
However light-cone superstring field theory suffers from contact term divergences
which have not yet been understood fully\cite{gr1,gr2,gr3,greenseiberg}.
A way to circumvent this has been suggested in \cite{1605.04666}.}

While string field theory is essential for
carrying out this computation to higher loop order, for one loop correction
to the masses one does not require the full power of string field theory. The reason 
has already been mentioned earlier: 
one loop mass renormalization can be computed from one loop two point
function of external states that satisfy tree level on-shell condition. 
No subtraction is necessary, unlike in the case of higher loop two point functions from
which the contribution from 1PR graphs have to be subtracted.
Nevertheless since this one loop two point function diverges due to the reasons
mentioned above, we need a way to deal with this divergence. 
The strategy we follow is to isolate the divergent part
and
reinterpret this as coming from a specific Feynman diagram of string field
theory. If we try to express this as integration over Schwinger parameters, we get
back the expression that we have in the world-sheet description, and it is
divergent. But we can 
directly evaluate this Feynman diagram by performing integration
over loop momenta following the prescription of \cite{1604.01783} 
and this yields a finite answer.
The difference between the two can be traced to the fact that the Schwinger parameter
representation of the internal propagators breaks down for certain range of momentum
integration. Since from the point of view of string field theory, the Feynman diagrams 
are more fundamental, the procedure of evaluating the Feynman diagrams directly
is the correct one, even when its Schwinger parameter representation fails.

The rest of the paper is organized as follows. In \S\ref{stoy} we 
introduce a toy quantum field theory that shares some essential properties
of string theory.   We compute one loop mass renormalization of an unstable particle
in this theory and show that we get a finite answer. On the other hand if we try to
evaluate the same expression by using Schwinger parameter representation of the
propagators, we get a divergent result. The divergence can be traced to the
breakdown of the Schwinger parameter representation of the propagator. In
\S\ref{sstring} we compute one loop mass renormalization of the lowest massive
string state of ten dimensional heterotic string theory on the leading Regge 
trajectory. The answer, expressed as an integral over the moduli space of a
torus with two punctures, has certain divergences from the boundary of the
moduli space. We isolate the divergent piece, and by comparing it with the
result of \S\ref{stoy} in the Schwinger parameter representation of the propagator,
identify the divergent piece as the contribution from a specific Feynman diagram
of string field theory. This Feynman diagram is then evaluated using direct momentum
space integration, leading to finite answer. Our final result is expressed as a sum
of three terms, given in \refb{edefJ1}, \refb{ei1spa} and 
\refb{ei2spa}, each of which is 
manifestly finite. 
We discuss extension of this analysis to general external states in
\S\ref{sgen} where we also give a justification of the procedure from string field
theory and show that the results for the renormalized mass obtained this way
agree for different versions of string field theory.
We also describe how our analysis can be easily extended to
compactified string theories. In \S\ref{sunitarity} we show
that the imaginary part of the mass$^2$ computed using our approach is
manifestly consistent with unitarity. In appendix \ref{sb} we show the equivalence between the 
$i\eps$ prescription of \cite{berera2,1307.5124} and our prescription of \S\ref{stoy} in the context of one
loop two point functions. In appendix \ref{sa} we analyze in detail
the `stringy
contribution' to mass renormalization given by \refb{edefJ1} 
and show explicitly that this gives a finite contribution.

\sectiono{Toy model} \label{stoy}

\begin{figure}
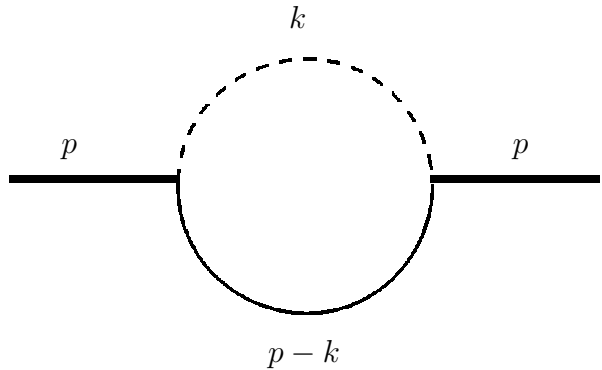


\begin{center}

\figqft

\vskip -25pt 

\caption{One loop mass renormalization diagram of a heavy state, labelled by a thick
line, due to a loop of light particles, labelled by thin lines. The dashed line
corresponds to a light particle of mass $m_1$
carrying momentum $k$ and the continuous thin line corresponds to a 
light particle of mass $m_2$ carrying momentum
$(p-k)$. All momenta flow from left
to right. \label{fqft}}

\end{center}

\end{figure}

Let us consider a quantum field theory in $D$ space-time dimensions
with three particles of masses $M$, $m_1$ and $m_2$ respectively, 
with $M>m_1+m_2$, in which there is a three point vertex that couples the three 
particles. Our goal will be to analyze the one loop mass
renormalization diagram shown in Fig.~\ref{fqft}. Inspired by string field
theory, we shall assume that the vertex 
contains a factor of $\exp[- {1\over 2} A \{k^2+m_1^2\} - {1\over 2}
A \{(p-k)^2 + m_2^2\}]$ for some positive constant $A$ that makes the diagram
ultraviolet (UV) finite\cite{1604.01783}. 
In that case the contribution of this diagram to mass$^2$ of the
heavy particle can be expressed as
\be \label{e1}
\delta M^2 = i\, B \, \int{d^D k \over (2\pi)^D} \, \exp[-A \{k^2+m_1^2\} -
A \{(p-k)^2 + m_2^2\}]\, \{k^2+m_1^2\}^{-1} \{(p-k)^2 + m_2^2\}^{-1}\, ,
\ee
where $B$ is another positive constant that includes multiplicative constant
contributions to the vertices, and $p$ is an on-shell external momentum
satisfying  $p^2=-M^2$. 
In general we could include factors involving 
polynomials in the momenta in the vertices without affecting the  UV finiteness, but
we have not included them to keep the analysis simple.  
Later we shall consider the effect of including such interactions. 

\subsection{Direct evaluation} \label{sdirect}

Using $k^2 = -(k^0)^2 + \vec k^2$ where $\vec k$ denotes $(D-1)$-dimensional
spatial momenta, we see that the exponential factor falls off exponentially as
$|\vec k|\to\infty$ but grows exponentially as $k^0\to \pm\infty$. This shows that
we cannot take the $k^0$ integral to run along the real axis. This issue was discussed
in detail in \cite{1604.01783} 
where we proposed that the ends of the $k^0$ integral must always
be at $\pm i\infty$ to ensure convergence of the integral, but the
integration contour may take
complicated form in the interior of the complex
$k^0$ plane to avoid poles of the propagator. This is done as follows:
begin with imaginary $p^0$ for which the $k^0$ contour is taken along the
imaginary axis and then deform $p^0$ to the physical real value staying in the first
quadrant of the complex $p^0$ plane, simultaneously deforming 
the $k^0$ contour appropriately to
always stay away from the poles. 
In particular \refb{e1} was analyzed in detail in \cite{1604.01783} 
using this prescription. Here
we shall review some of the important details of that analysis.

The integrand of \refb{e1} has poles in the $k^0$
plane at
\be  \label{e2}
Q_1 \equiv \sqrt {\vec k^2 + m_1^2}, \quad Q_2 \equiv -\sqrt{\vec  k^2 + m_1^2}, \quad
Q_3 \equiv p^0 + \sqrt{(\vec p - \vec  k)^2 + m_2^2} , \quad Q_4 \equiv
p^0 - \sqrt{(\vec p - \vec  k)^2 + m_2^2}\, .
\ee
For imaginary $p^0$, and $k^0$ contour running along the imaginary axis from
$-i\infty$ to $i\infty$, 
the poles $Q_1$ and $Q_3$ are to the right of the integration
contour whereas the poles $Q_2$ and $Q_4$ are to the left of the integration
contour. When $p^0$ is continued to the real axis along the first quadrant, the contour
needs to be deformed appropriately so that $Q_1$ and $Q_3$ continue to lie
on the right and $Q_2$ and $Q_4$ continue to lie on the left. There
are different possible configurations depending on the value of $\vec k$.

\begin{figure}
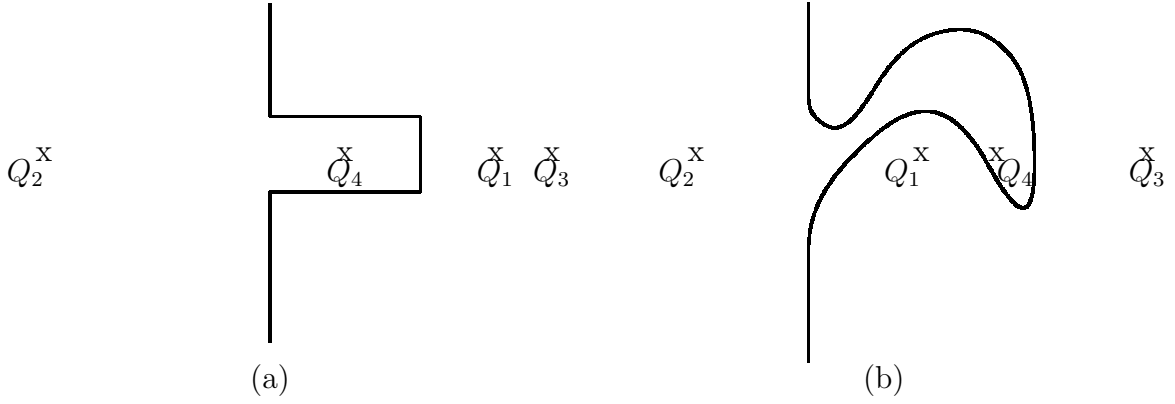


\begin{center}

\hbox{\figtwo \hfill \figfour}

\vskip 25pt 

\caption{The integrations contours in the $k^0$ plane.
\label{f2}}

\end{center}

\end{figure}

As long as $p^0< \sqrt{\vec  k^2+m_1^2} + 
\sqrt{(\vec p-\vec  k)^2+m_2^2}$,  
$Q_4$ lies to the left of $Q_1$ and the contour can be taken 
as shown in Fig.~\ref{f2}(a). 
On the other hand for $p^0 > \sqrt{\vec  k^2+m_1^2} + 
\sqrt{(\vec p-\vec  k)^2+m_2^2}$, $Q_4$ is to the right of $Q_1$ and the 
deformed contour takes the form shown in Fig.~\ref{f2}(b). In drawing this
we have used the fact that when $p^0$ lies in the first quadrant,
$Q_4$ remains above $Q_1$ as it 
passes $Q_1$ and that during this process the contour needs to be
deformed continuously without passing through a pole. At the
boundary between these two regions $Q_4$ approaches
$Q_1$. In this case we have to use a limiting procedure to determine the contour, and the
correct procedure will be to take $p^0$ in the first quadrant, evaluate the integral
and then take the limit of real $p^0$. This in particular means that $Q_4$ approaches
$Q_1$ from above in this limit.

In order to evaluate the integral, in both cases we deform the $k^0$ contour to be a 
sum of a contour along the imaginary axis and an anti-clockwise contour around the
pole at $Q_4$. 
We shall choose, for convenience,
\be \label{eonshell}
p=(M, \vec 0)\, .
\ee
In this case the contribution from the first contour, after relabeling $k^0$ as $i\, u$,
takes the form
\ben \label{ei1}
I_1&=&-B \int{d^{D-1}k\over (2\pi)^{D-1} }\, \int_{-\infty}^\infty {du\over 2\pi} \,
\exp\left[- A\left\{u^2 +\vec k^2+m_1^2\right\} - A\left\{(u+iM)^2 
+ \vec k^2+m_2^2\right\}\right] \nonumber \\
&& \left(u^2 + \vec k^2 + m_1^2\right)^{-1}
\left\{(u+iM)^2 + \vec k^2+m_2^2\right\}^{-1}\, .
\een
On the other hand the
contribution from the residue at $Q_4$ gives
\ben \label{ei2-}
I_2 &=& -B  \int{d^{D-1}k\over (2\pi)^{D-1} }
\exp\left[A \left(M - \sqrt{\vec k^2+m_2^2}\right)^2 - A (\vec k^2 + m_1^2)
\right] \Theta \left(M - \sqrt{\vec k^2+m_2^2}\right)
\nonumber \\ && 
\left(2  \sqrt{\vec k^2 + m_2^2} \right)^{-1}
\left\{ M+\sqrt{\vec k^2 +m_1^2} - \sqrt{\vec k^2+m_2^2}\right\}^{-1}
\nonumber \\ 
&& \left\{ \sqrt{\vec k^2 + m_1^2} + \sqrt{\vec k^2+m_2^2} - M-i\eps\right\}^{-1}\, .
\een
In this expression $\Theta$ denotes the Heaviside function and reflects that this
contribution is present only when $Q_4$ is to the right of the imaginary axis.
The $i\eps$ in the arguments of the last term represents that we need to take the
limit $p^0\to M$ from the first quadrant, i.e. set $p^0$ to $M+i\eps$ and then take the
$\eps\to 0^+$ limit.  Defining $v=|\vec k|$ and doing the angular integration, $I_1$ and
$I_2$ may
be rewritten as
\ben \label{ei1fin}
I_1&=&-B \, (2\pi)^{-D} \Omega_{D-2} 
\,  \int_0^\infty  dv\, \, \int_{-\infty}^\infty {du} \, v^{D-2}\, 
\exp\bigg[- A\left\{u^2 +v^2+m_1^2\right\} \nonumber \\
&& - A\left\{(u+iM)^2 
+ v^2+m_2^2\right\}\bigg] \left(u^2 + v^2 + m_1^2\right)^{-1}
\left\{(u+iM)^2 + v^2+m_2^2\right\}^{-1}\, ,
\een
and
\ben \label{ei2fin}
I_2 &=& -B  \, (2\pi)^{-(D-1)} \Omega_{D-2} \int_0^{\sqrt{M^2-m_2^2}} \, dv\, v^{D-2}\, 
\exp\left[A \left(M - \sqrt{v^2+m_2^2}\right)^2 - A (v^2 + m_1^2)
\right]\nonumber \\
&& \left(2\sqrt{v^2 + m_2^2}\right)^{-1} 
\left\{ M + \sqrt{v^2+m_1^2} - \sqrt{v^2+m_2^2}\right\}^{-1} \nonumber \\ &&
\left\{ \sqrt{v^2 + m_1^2} + \sqrt{v^2 + m_2^2} - M-i\eps\right\}^{-1}\, ,
\een
where $\Omega_{D-2}$ is the volume of the unit $(D-2)$ sphere.
Due to the exponential suppression factors and/or limits of integration,
neither $I_1$ nor $I_2$ has any divergence from the large $u$ or large $v$ region.
Even though as $\epsilon\to 0$
the integrand of $I_2$
has a pole on the real $v$ axis from the last term,
the contour is not pinched there. Hence we can define the integral by deforming the
$v$ integration contour below the real axis, getting a finite result. Therefore 
both $I_1$
and $I_2$ are manifestly finite (and in particular can be evaluated using numerical
integration).

The analysis given above can be easily generalized to the case where the
integrand in \refb{e1} is multiplied by an additional polynomial 
in momenta
coming from the vertices and/or the propagators. 
Using rotational invariance we can always replace this by a polynomial $Q$
in $k^0$ and $\vec k^2$.
The result will still be given
by the sum of two terms like \refb{ei1fin} and \refb{ei2fin}. The integrand
in \refb{ei1fin} will now be multiplied by the polynomial $Q$ with $k^0$ replaced by
$i\, u$ and $\vec k^2$ replaced by $v^2$. On the other hand the integrand in
\refb{ei2fin} will be multiplied by the polynomial $Q$ with $k^0$ replaced by
$M-\sqrt{v^2+m_2^2}$ and $\vec k^2$ replaced by $v^2$. 

\subsection{Schwinger parameter representation} \label{ssch}

We shall now try to
evaluate \refb{e1}  by representing the propagators as integrals
over Schwinger parameters. For this we write
\ben \label{esp1}
(k^2 + m_1^2)^{-1} &=& \int_0^\infty ds_1 \, \exp\left[-s_1(k^2+m_1^2)\right],
\nonumber \\ 
\{(p-k)^2 + m_2^2\}^{-1} &=& \int_0^\infty ds_2 \, \exp\left[-s_2\{(p-k)^2+m_2^2\}\right],
\een
and substitute into \refb{e1}. This give
\be \label{eddm}
\delta M^2 
= i\, B \,\int_0^\infty  ds_1 \int_0^\infty ds_2 \,  \int{d^D k \over (2\pi)^D} \, 
\exp\left[ -(A+s_1) \{k^2+m_1^2\} -
(A+s_2) \{(p-k)^2 + m_2^2\}\right]\, .
\ee 
After performing integral over $k$, pretending that the $k^0$ integral runs along the
imaginary axis and is convergent, and defining new variables
\be 
t_1 = s_1+A, \quad t_2 = s_2+A\, ,
\ee
we get
\be \label{efinint}
\delta M^2 = - B \, (4\pi)^{-D/2} \, \int_A^\infty  dt_1 \int_A^\infty dt_2 \, 
(t_1+t_2)^{-D/2} \, 
\exp\left[ {t_1 t_2\over t_1+t_2}\, M^2 - (t_1 m_1^2+t_2 m_2^2)\right]\, .
\ee
This expression has no UV divergence, i.e. divergence from the small $t_i$
region, since the lower limits of $t_i$ integrals are shifted to positive values $A$.
However it is easy to see that this integral diverges from the region $t_1,t_2\to\infty$
if
\be
M>m_1+m_2\, .
\ee
This divergence can be traced to the fact that for $M>m_1+m_2$, it is not possible to choose
the $k^0$ integration contour in a way that keeps the real parts of
both $k^2+m_1^2$ and
$(p-k)^2 + m_2^2$ positive. As a result the Schwinger parameter representation
\refb{esp1} breaks down. However note that we can get finite results by
taking the
upper limits of the $t_i$ integrals to be $i\infty$ 
instead of $\infty$\cite{berera2,1307.5124}.
We have shown in appendix \ref{sb} that this gives the same result as what we would obtain by
following the
prescription of \S\ref{sdirect} for evaluating \refb{e1}.

Since string world-sheet description of the S-matrix elements naturally gives the
amplitudes in the Schwinger parameter representation, we shall see that the world-sheet
description of one loop mass renormalization in string theory encounters similar 
divergences. Our strategy will be to 
use the insight gain from our analysis above to convert this
to a momentum space integral of the form given in \refb{e1}
and extract finite answers.
For this we shall need a generalization of the analysis given above, 
where the integrand in \refb{e1} has an additional multiplicative factor given by
some polynomial in the momenta $\{k^\mu\}$. We shall first discuss a few 
examples. The first example we consider is when the integrand in \refb{e1}
has an additional factor of
$(k^0)^2$. In this case  it is easy to see that the integrand in \refb{efinint} will be
multiplied by an additional factor of
\be 
-{1\over 2(t_1+t_2)} + {t_2^2 \over (t_1+t_2)^2} M^2 \, .
\ee
Next we consider the case where the integrand in \refb{e1}
has a multiplicative factor of
$k^0$. In this case the integrand in \refb{efinint} is multiplied by an additional
factor of 
\be
{t_2\over t_1+t_2} M\, .
\ee
If we consider the case where the integrand in \refb{e1} has an additional 
multiplicative factor of $k^i k^j$ with $1\le i,j\le (D-1)$, then we get an
additional multiplicative factor of
\be\label{eaddmu0}
\delta_{ij}\, {1\over 2(t_1+t_2)}
\ee
in \refb{efinint}.
Finally if the integrand has an additional factor of $k^i k^j k^m k^n$ then we
get an additional multiplicative factor of
\be \label{eaddmul}
{1\over 4 (t_1+t_2)^2} \left(\delta_{ij} \delta_{mn} + \delta_{im} \delta_{jn}
+ \delta_{in} \delta_{jm}\right)\, .
\ee

It is clear that given any polynomial in $\{k^\mu\}$ inserted into \refb{eddm},
we can find the corresponding
insertion in the integrand of the Schwinger parameter representation 
\refb{efinint} by formally
carrying out the integration over momenta using the rules of gaussian integration,
pretending that the integral is convergent. An interesting question is whether the
reverse is true: given any polynomial $P$ in $1/(t_1+t_2)$ and $t_2/(t_1+t_2)$, can
we find a function $Q$ of momenta such that the following holds?
\ben \label{exyx}
&& i \int{d^D k \over (2\pi)^D} \, 
\exp\left[ - t_1 \{k^2+m_1^2\} -
t_2 \{(p-k)^2 + m_2^2\}\right] Q(k) \nonumber \\
&=& - (4\pi)^{-D/2} 
\exp\left[ {t_1 t_2\over t_1+t_2}\, M^2 - (t_1 m_1^2+t_2 m_2^2)\right] P
\left({1\over t_1+t_2}, {t_2\over t_1+t_2}\right)\, .
\een
It is clear that due to rotational invariance of the
problem $Q$ cannot be unique -- {\it e.g.} $(k^1)^2$, $(k^2)^2$ and
$\vec k^2 / (D-1)$ will all generate the same expression after momentum integration.
However they will also give the same result if we insert $Q(k)$ into the integrand
in \refb{e1} and carry out the momentum integration
directly using the procedure described in \S\ref{sdirect}.
Therefore 
we can easily resolve this ambiguity in the form of $Q$
by restricting $Q$ to be a polynomial in
$k^0$ and $\vec k^2$. In that case we can construct a unique $Q$ from a given $P$
as follows.
We can start from the terms in $P$ 
with the highest power of $t_2/(t_1+t_2)$, and among these
the term with highest power of $1/(t_1+t_2)$. If this has the form
$\{t_2/ (t_1+t_2)\}^n \{1/(t_1+t_2)\}^m$, then we need a term $Q_1$ 
in $Q$ proportional to
$(k^0)^n (\vec k^2)^m$ to generate this. Let $P_1$ be the polynomial in $1/(t_1+t_2)$
and $t_2/(t_1+t_2)$ 
obtained by replacing $Q,P$ by $Q_1,P_1$ in \refb{exyx}.
Besides containing the term proportional to
$\{t_2/ (t_1+t_2)\}^n \{1/(t_1+t_2)\}^m$ appearing in $P$, 
$P_1$ will generically also contain terms
with lower powers of $t_2/ (t_1+t_2)$. 
We now repeat the analysis for $P-P_1$, by identifying the terms in $P-P_1$ with
highest power of $t_2/(t_1+t_2)$, and among them the term with highest power
of $1/(t_1+t_2)$. Proceeding this
way till we have exhausted all the terms in $P$,
we can find the polynomial $Q=Q_1+Q_2+\cdots$ that, when inserted into 
the left hand side of \refb{exyx}, will produce the desired $P$ on the right hand side.

The effect of inserting $P$ in the integrand of \refb{efinint} can now be represented by 
insertion of $Q(k)$ in the integrand of \refb{eddm} and hence of \refb{e1}.
Since $Q$ is a polynomial in $\{k^\mu\}$, there will be
no difficulty in carrying out the momentum integration in \refb{e1}
directly following
the procedure described in \S\ref{sdirect} to get a finite result.
This way any integral of the form \refb{efinint}, with arbitrary polynomial
of $1/(t_1+t_2)$ and $t_2/(t_1+t_2)$ inserted in the integrand, can be
interpreted as a finite momentum space integral.

\sectiono{One loop mass renormalization of an unstable state in string theory} 
\label{sstring}

We shall now use the insight gained from the analysis of \S\ref{stoy} to compute
one loop mass renormalization in string theory. 
In this section we shall consider a specific example, leaving the general analysis to
\S\ref{sgen}.
We  consider the lowest massive state on the leading Regge 
trajectory
in the SO(32) heterotic string
theory.\footnote{The advantage of working with states 
on the leading Regge trajectory
is that they do not mix with any other 
state at the same mass level. This simplifies our analysis, but the method that we
shall describe is valid for arbitrary states.} 
The one loop correction to the mass$^2$ of this state
can be computed from the on-shell two point function of the corresponding vertex
operators on the torus.
If we define  
\be 
X^\pm = (X^1 \pm i X^2), \quad \psi^\pm = (\psi^1 \pm i \psi^2)\, ,
\ee
where $X^\mu$ are the world-sheet scalars and $\psi^\mu$ are the right-moving
world-sheet fermions, then the $-1$ picture unintegrated
vertex operators of the states whose
two point function on the torus we need to compute are:
\be
\bar c\, c\, e^{-\phi} \psi^+ \p X^+ (\bar \p X^+)^2 e^{i k^0 X^0}\quad \hbox{and} \quad 
\bar c\, c\, e^{-\phi} \psi^- \p X^- (\bar \p X^-)^2 e^{-i k^0 X^0}\, ,
\ee 
up to overall normalization constants. Here $\phi$ is the world-sheet scalar that
originates from bosonizing the $\beta$-$\gamma$ ghost system, and $c$, $\bar c$ are
the usual ghost fields associated with diffeomorphism invariance on the world-sheet.
Now it was argued in \cite{1304.0458} that all the states
at the first massive level which differ from each other by
different right-moving excitations are related by space-time supersymmetry and hence
will have the same mass renormalization. Using this we can instead consider the
vertex operators
\be \label{enewv}
\bar c\, c\, e^{-\phi} \psi^1 \psi^2 \psi^3 (\bar \p X^+)^2 e^{i k^0 X^0}\quad \hbox{and} \quad 
\bar c\, c\, e^{-\phi} \psi^1 \psi^2 \psi^3 (\bar \p X^-)^2 e^{-i k^0 X^0}\, .
\ee 
The reason for doing this is that with this choice the right-moving parts of the
vertex operators become identical to those used in \cite{1304.0458} and we can make use
of the results of \cite{1304.0458}.\footnote{In 
\cite{1304.0458} we converted both vertex operators to
zero picture vertex operators for carrying out the computation. This does not satisfy
the correct factorization condition when two vertex operators approach each other,
and in some cases, can give erroneous results\cite{1209.5461,1408.0571}. 
However for flat space-time
background, including toroidal compactification, the difference between the correct
result and the one obtained using zero picture vertex operators can be computed
using the analysis given in \cite{1408.0571} and can be shown to vanish.} 
In this case the only difference between the vertex operators used in \cite{1304.0458} and
those used here is that the left-moving part of the vertex operators used in 
\cite{1304.0458} were $\bar S_\alpha$ -- the spin fields of the left-moving world-sheet 
fermions responsible for the SO(32) gauge group --
instead of $(\bar\p X^\pm)^2$. Therefore if we want to compute
the two point correlation function of the vertex operators \refb{enewv} inserted at
0 and $z$ on a torus with modular parameter $\tau$, all we need to do is to replace,
in the result of \cite{1304.0458}, the normalized two point function
$\langle \bar S_\alpha(\bar z) \bar S_\beta(0)\rangle$ by the normalized two point
function $\langle (\bar \p X^+(\bar z))^2 (\bar \p X^-(0))^2\rangle$. Normalizing both
correlators so that as $\bar z\to 0$ they go as $1/\bar z^4$, we have
\be \label{eone}
\langle \bar S_\alpha(\bar z) \bar S_\beta(0)\rangle 
=\delta_{\alpha\beta}  \left(\sum_\nu \overline{{\vt_\nu(z/2)}}^{16}\right)
\left( \sum_\nu \overline{\vt_\nu(0)}^{16}\right)^{-1} \left(\overline{\vt_1'(0)}\right)^{4}
\left(\overline{\vt_1(z)}\right)^{-4}\, ,
\ee
and
\be \label{etwo}
\langle  (\bar \p X^+(\bar z))^2 (\bar \p X^-(0))^2\rangle
= \left[\left( {\overline{\vt_1'(z)}\over \overline{\vt_1(z)}}\right)^2 
- {\overline{\vt_1''(z)}\over \overline{\vt_1(z)}} - {\pi\over \tau_2}\right]^2\, ,
\ee
where 
$\vt_\nu$ for $1\le\nu\le 4$ 
denotes Jacobi theta function of spin structure $\nu$, with $\vt_1$
being the Jacobi theta function with odd spin structure, and
$\tau_1,\tau_2,z_1,z_2$ are defined via
\be 
\tau=\tau_1+i\tau_2, \qquad z=z_1+i z_2\, .
\ee
Therefore, to compute $\delta M^2$ we have to multiply the integrand obtained in
\cite{1304.0458} by the ratio of \refb{etwo} and \refb{eone}. This gives, from 
eq.(4.16), (4.17)  of \cite{1304.0458}:
\ben \label{edelM}
\delta M^2  &=& -{1\over 32 \, \pi} \, M^2 \, g^2\, 
\int d^2 \tau \int d^2 z\,  F(z,\bar z, \tau, \bar\tau) \, , \nonumber \\
F(z,\bar z, \tau, \bar\tau) &\equiv & \left\{\sum_{\nu} \overline{\vt_{\nu}(0)^{16}}\right\}
(\overline{\eta(\tau)})^{-18} (\eta(\tau))^{-6} (\overline{\vt_1'(0)})^{-4}
\left( {\vt_{1}(z)\overline{\vt_{1}(z)}}\right)^2 \nonumber \\ &&
\left[\left( {\overline{\vt_1'(z)}\over \overline{\vt_1(z)}}\right)^2 
- {\overline{\vt_1''(z)}\over \overline{\vt_1(z)}} - {\pi\over \tau_2}\right]^2
\exp[-{4\pi \, z_2^2 / \tau_2}] \, (\tau_2)^{-5}\, ,
\een
where $g$ is the string coupling constant, normalized as in \cite{1304.0458}.
The integration over $\tau$ runs over the fundamental region and that over $z$
runs over the whole torus. The sum over $\nu$ in \refb{edelM} comes from the sum over
spin structures in the left-moving sector of the world-sheet -- the sum over spin structures
in the right-moving sector have already been performed\cite{1304.0458} in arriving
at \refb{edelM}. Analogous expression for arbitrary state
on the leading Regge trajectory in type II string theory can be found in
\cite{sundborg,0210245}.

We shall now try to analyze possible divergences in this integral. It is easy to
see that the integral has no divergence from the $z\to 0$ region, and is in fact finite for all
finite values of $\tau$ and $z$. Since $z_1$ and $\tau_1$ integrals are 
restricted to the range (0,1) and the $z_2$ integral is restricted to the range
$0\le z_2<\tau_2$, possible
divergences come from the region of large $\tau_2$ and possibly large $z_2$.
In particular we can remove  the $\tau_2<1$ region from our consideration, 
since this is a finite region with
bounded integrand.
For $\tau_2\ge 1$ the $\tau_1$ and $z_1$ 
integrals
run over the entire range between 0 and 1. 
While evaluating these integrals
we need to first 
integrate over $z_1$ and $\tau_1$ for fixed $z_2$ and $\tau_2$, and then
integrate over $z_2$ and $\tau_2$. A justification for this from string
field theory will be
given in \S\ref{sgen}. 
Therefore if we expand the integrand in this region
in powers of $e^{2\pi i \tau}$,
$e^{-2\pi i \bar\tau}$, $e^{2\pi i z}$
and $e^{-2\pi i \bar z}$, all terms with non-zero powers of $e^{2\pi i \tau_1}$
or $e^{2\pi i z_1}$ will integrate to zero, and only the $\tau_1$ and $z_1$
independent terms will survive.

We shall first consider the 
large $\tau_2$ but finite $z_2$ region. For this we define finite $z_2$ region
to be the region $z_2< \Lambda$ for some fixed positive number $\Lambda\le\tau_2$.
In this region
$F(z,\bar z, \tau, \bar\tau) $
has the form
\be \label{efflimit}
F(z,\bar z, \tau, \bar\tau)=
\exp[-{4\pi \, z_2^2 / \tau_2}] \, (\tau_2)^{-5} \left[
 2\, \pi^{-4} \, e^{2\pi i \bar\tau}|\sin(\pi z)|^4 \left( \pi^2 \cot^2(\pi 
\bar z) + \pi^2 -{\pi\over \tau_2}\right)^2
+\OO(1)\right]\, ,
\ee
where
the $\OO(1)$ term is finite for any finite $z,\tau$ and 
approaches a fixed finite function of $z$ for $\tau_2\to\infty$ and finite 
$z$.\footnote{For $z=0$ the $\OO(1)$ term has a phase ambiguity since both the
function  $F$ and the first term inside the square bracket in \refb{efflimit}
is proportional to $(z/\bar z)^2$ for small $z$. 
But the integral of this term over any finite neighbourhood of $z=0$ 
is unambiguous
and finite for all $\tau$ inside the fundamental domain, as well as
in the $\tau\to i\infty$
limit. 
The analogous expression in type II string theory will not have any such
phase ambiguity.}
Therefore
for $z_2<\Lambda$, the $\OO(1)$ term  inside the square bracket
can be bounded from above by a positive number $\Delta$, and
after integration
over $z$ and $\tau$ restricted to the region $z_2<\Lambda\le\tau_2$, $\tau_2\ge 1$,
its contribution to $\int d^2\tau d^2 z\, F$ will be bounded from above by
\be
\Lambda\, \Delta\, \int_1^\infty d\tau_2 \, \tau_2^{-5} =\Lambda \, \Delta / 4\, .
\ee
On the other hand
the term proportional to $e^{2\pi i\bar\tau}$ inside the square bracket 
in \refb{efflimit} gives
vanishing contribution after the  $\tau_1$ integration.
This shows that the integral does not receive any divergent contribution from the
$z_2<\Lambda$ and large $\tau_2$ region.

Next we examine the region of integration where both $\tau_2$ and $z_2$ are large.
Note that due to the reflection symmetry $z\to \tau-z$ of the integrand, there is
also no divergence from the region where $\tau_2$ and $z_2$ are large with 
$\tau_2-z_2$ finite; so we focus on the region where $z_2$ and $\tau_2-z_2$ are
both large. 
Expanding the integrand in powers of $e^{2\pi i \tau}$, $e^{2\pi i z}$ and their complex
conjugates, and throwing away all terms which have non-zero
powers of $e^{2\pi i z_1}$ and/or $e^{2\pi i \tau_1}$ since they vanish after
integration over $z_1$ and $\tau_1$, we find that the part of 
$F(z,\bar z)$ that can give divergent contribution to \refb{edelM} takes the form
\be\label{ediv}
2\, (2\pi)^{-4} \, \left( 32 \pi^4 - 32 {\pi^3\over \tau_2} +
512 \, {\pi^2\over \tau_2^2} \right) \,
\exp[4\pi z_2 - 4\pi z_2^2/\tau_2]\, \tau_2^{-5}\, .
\ee

Based on the above understanding of the possible sources of divergence, we shall now 
give a systematic procedure for isolating and dealing with the potentially divergent
part.
Using \refb{edelM} we can write
\be \label{efindelM}
\delta M^2 = J_1+J_2\, ,
\ee
where 
\ben \label{edefJ1}
J_1 &=& -{1\over 32 \, \pi} \, M^2 \, g^2 
\int d^2 \tau \int d^2 z\, \bigg[F(z,\bar z, \tau, \bar\tau) \\ 
&& 
 -   \Theta(\tau_2 - z_2 - \Lambda) \Theta(z_2-\Lambda) \,
 2\,  (2\pi)^{-4} \, \left( 32 \pi^4 -32 {\pi^3 \over \tau_2} +
512 \, {\pi^2\over\tau_2^2}\right)  \,
\exp[4\pi z_2 - 4\pi z_2^2/\tau_2] \, \tau_2^{-5}
\bigg] \, ,
\nonumber
\een
and
\ben \label{edefJ2}
J_2 &=& -{1\over 32 \, \pi} \, M^2 \, g^2
\int d^2 \tau \int d^2 z\, 
2\,  (2\pi)^{-4} \, \left( 32 \pi^4 -32 {\pi^3 \over \tau_2} +
512 \, {\pi^2\over\tau_2^2}\right)  \,
\exp[4\pi z_2 - 4\pi z_2^2/\tau_2] \, \tau_2^{-5}
\nonumber \\
&&  \hskip 2in \Theta(\tau_2 - z_2 - \Lambda) \, \Theta(z_2-\Lambda)\, ,
\een
where $\Lambda$ is an arbitrary positive constant larger than 1,
and $\Theta$ denotes
Heaviside step function. 

First let us analyze $J_1$. For this it will be convenient to 
define the variable
\be 
w = \tau-z=w_1+i w_2\, ,
\ee
and
divide the integration region into four parts.
The region $z_2< \Lambda$, $w_2< \Lambda$ has finite size, and the integrand
$F$ is bounded. Hence there is no divergence from this region. In the region 
$z_2 < \Lambda$, $w_2\ge \Lambda$ the integrand is $F$ and by our previous
argument that there is no divergence from the finite $z_2$, large $\tau_2$ region,
this integral is also finite. The region $z_2\ge\Lambda$, $w_2<\Lambda$ is related
to the one just described by the $z\leftrightarrow w$, or equivalently $z\to\tau-z$ symmetry,
and gives finite result. This leaves us with the region
$z_2\ge\Lambda$, $w_2\ge \Lambda$. In this region the term proportional to the
Heaviside functions
in \refb{edefJ1} subtracts the leading divergent piece.
A careful analysis (see appendix \ref{sa}) shows that
after throwing away all
terms carrying non-zero powers of $e^{2\pi i z_1}$ and $e^{2\pi i w_1}$, we get
finite result for $J_1$ from the $z_2\ge\Lambda$, $w_2\ge \Lambda$ region.
Therefore there are no divergences in $J_1$ from any part of the region of
integration.

Next we turn to the analysis of $J_2$ which only receives contribution from the
$z_2\ge\Lambda$, $w_2\ge\Lambda$ region. 
We can bring $J_2$ to a more recognizable form by performing integrations over
$z_1$ and $\tau_1$ and defining the variables
\be
t_1 = \pi \, z_2, \qquad t_2 = \pi w_2 = \pi (\tau_2-z_2)\, .
\ee
In terms of these variables $J_2$ takes the form
\be \label{enewJ2}
J_2 = - 2^{-3} \pi^{2} \, M^2\, g^2  \int_{\pi\Lambda}^\infty dt_1 \int_{\pi\Lambda}^\infty dt_2\, 
(t_1+t_2)^{-5} \,  \exp\left[4 {t_1 t_2\over t_1+t_2}\right] \,
\left\{ 1 - {1 \over (t_1+t_2)} +
16 \, {1\over (t_1+t_2)^2}\right\}\, .
\ee 
The integral has apparent divergence from the large $t_1,t_2$ region. However we
shall now try to
interpret it as a finite momentum space integral by
comparing this with \refb{efinint} for $D=10$.
Comparing the overall normalization and the argument of the exponential 
we get\footnote{The peculiar factor of $(2\pi)^7$ in the expression for $B$ can be
traced to the fact that the heterotic string coupling $g_H$ is related to the
coupling $g$ used here by the relation $g_H = (2\pi)^{7/2} g $\cite{1304.0458}.}
\be\label{epar}
B = (2\pi)^{7} M^2 g^2, \quad A=\pi\Lambda, \quad M=2, \quad m_1=0, \quad m_2=0\, .
\ee
Matching the rest of the integrand in \refb{enewJ2} with 
what appears in
\refb{efinint} for $D=10$, we see that we have an extra insertion of a factor of
\be
\left( 1 - {1 \over (t_1+t_2)} +
16 \, {1\over (t_1+t_2)^2}\right)\, .
\ee
Using \refb{eaddmu0}, \refb{eaddmul} this can be identified as the 
effect of inserting 
a factor of 
\be \label{emom}
(1 -2 \, (k^1)^2 + 64 \, (k^1)^2 (k^2)^2)\, .
\ee
in the integrand in the momentum 
space.\footnote{Note that knowing the integrand in the Schwinger parameter
representation does not fix the form in momentum space completely, {\it e.g.} 
the multiplicative factor could also have been $(1 - (k^2)^2 +
64 (k^3)^2 (k^4)^2)$, or averages
of various factors of this form. If we had started from string field theory, then
Feynman diagrams would lead to a specific form. However for evaluation of the
integral the detailed form is not necessary since due to rotation symmetry
all of them lead to the same
value of the integral.}
Combining this with \refb{e1}, we can express $J_2$ as a momentum space
integral
\ben \label{e1aa}
J_2 &=& i\, (2\pi)^{7} M^2 g^2 \, \int{d^{10} k \over (2\pi)^{10}} \, 
\exp[-\pi\Lambda k^2 -
\pi\Lambda (p-k)^2 ]\, (k^2)^{-1} \{(p-k)^2\}^{-1} \nonumber \\ && \hskip 2in
\{1 -2 \, (k^1)^2 + 64 \, (k^1)^2 (k^2)^2\}  \, .
\een
This of course gives a finite contribution and can be evaluated using the method
described in \S\ref{sdirect}.

Therefore we see that $J_2$ can be identified as the contribution from the Feynman
diagram of the form shown in Fig.~\ref{fqft} with the parameters given in
\refb{epar}, and extra momentum dependent insertion in the integrand given in
\refb{emom}.
In the $\alpha'=1$ unit that we have been working in, $M=2$ is the correct mass
of the external state. 
The result
$m_1=m_2=0$ in \refb{epar} indicates that for this state the only source of divergence
comes from the graphs where the intermediate states are massless. 
$J_1$ can be regarded as the contribution from the Feynman
diagrams of Fig.~\ref{fqft} with other massive string states propagating in the
loop and from other
Feynman diagrams, including the elementary two point vertex. Note the dependence
of $J_1$ and $J_2$ on the arbitrary parameter 
$\Lambda$; this represents the freedom of
changing the interaction vertices of string field theory by `adding stubs', and 
can be compensated for by a redefinition of the string fields\cite{9301097}. We shall
show later that $J_1+J_2$ is independent of $\Lambda$.

Manipulating \refb{e1aa} as in \S\ref{sdirect} with $m_1=m_2=0$,
we can express this as
\be \label{ej3i1i2}
J_2 = I_1 + I_2\, ,
\ee
where
\ben \label{ei1sp}
I_1&=&- (2\pi)^{7} M^2 g^2  \int{d^{9}k\over (2\pi)^{9} }\, \int_{-\infty}^\infty {du\over 2\pi} \,
\exp\bigg[-  \pi\Lambda\left\{u^2 +\vec k^2\right\} 
-  \pi\Lambda\left\{(u+iM)^2 
+ \vec k^2\right\}\bigg] \nonumber \\
&& \times \, \{1 -2 \, (k^1)^2 + 64 \, (k^1)^2 (k^2)^2\} 
\,\left(u^2 + \vec k^2\right)^{-1}
\left\{(u+iM)^2 + \vec k^2\right\}^{-1}\, ,
\een
and
\ben \label{ei2sp}
I_2 &=& -(2\pi)^{7}M^2 g^2   \int{d^{9}k\over (2\pi)^{9} }
\exp\left[ \pi\Lambda \left(M - |\vec k|\right)^2 
-  \pi\Lambda \, \vec k^2 
\right] \, \Theta \left(M - |\vec k|\right) \nonumber \\ 
&&\times \,
\{1 -2 \, (k^1)^2 + 64 \, (k^1)^2 (k^2)^2\} \,
\left(2 M |\vec k|\right)^{-1}
\left\{ 2 |\vec k|- M-i\eps\right\}^{-1}\, . 
\een
We can simplify both expressions by noting that due to rotational invariance
the insertions of  $k^i k^j$ and 
$k^i k^j k^m k^n$ must give contributions proportional to
\be
\delta_{ij} \quad \hbox{and} \quad \delta_{ij} \delta_{mn} + \delta_{im} \delta_{jn}
+ \delta_{in} \delta_{jm}\, ,
\ee
respectively.
This allows us to replace the insertion of $(k^1)^2$ by $\vec k^2/9$ and
$(k^1)^2 (k^2)^2$ by
$(\vec k^2)^2/99$. Defining $v=|\vec k|$ we can write
\ben \label{ei1spa}
I_1&=&-
(2\pi)^{-3} M^2 g^2 \, {\Omega_8}\, 
\int_0^\infty dv \, \int_{-\infty}^\infty {du} \, v^8\, 
\exp\left[-  \pi\Lambda\left\{u^2 +v^2\right\} 
-  \pi\Lambda\left\{(u+iM)^2 
+ v^2\right\}\right] \nonumber \\ && 
\hskip .5in \left(1 - {2\over 9} \, v^2 + {64\over 99} \, v^4\right)  \,
\left(u^2 + v^2\right)^{-1}
\left\{(u+iM)^2 + v^2\right\}^{-1}\, .
\een
On the other hand 
$I_2$ takes the form:
\ben \label{ei2spa}
I_2 &=& -(2 \pi)^{-2}M^2 g^2  \, {\Omega_8} \,
\int_0^M dv 
\, v^8 \, 
\exp\left[ \pi\Lambda \left(M - v\right)^2 
-  \pi\Lambda \, v^2 
\right]  \nonumber \\&&
\hskip 1in \left(1 - {2\over 9} \, v^2 + {64\over 99} \, v^4\right)  \,
\left(2 M v\right)^{-1}
\left\{ 2 v - M-i\eps\right\}^{-1}\, . 
\een
$I_1$ is manifestly finite. $I_2$ is also manifestly finite if we deform the integration 
contour to avoid the pole at $v=(M+i\eps)/2$
by taking it to lie below the real axis. This gives a 
completely finite result for $\delta M^2$, given by the sum of $J_1$,
$I_1$ and $I_2$.  

It is easy to see that $I_1$ is real. We can also see the reality of $J_1$
given in \refb{edefJ1} by
observing that
\be
( F(z,\bar z, \tau, \bar\tau))^*= F(-\bar z,- z, -\bar\tau, -\tau)\, ,
\ee
and that the integration domain and the integration measure are invariant
under $(z\leftrightarrow -\bar z, \tau\leftrightarrow -\bar\tau)$.
Therefore the imaginary part of the amplitude comes only from
$I_2$. This  
can be isolated by replacing the last factor in \refb{ei2spa} by a
sum of the principal value and a delta function
and noting that the imaginary part comes from the delta function. This gives 
\ben 
{\rm Im} \, \left(\delta M^2\right) &=&
- {1\over 4\pi} M^2 g^2   {\Omega_8} 
\int_0^M dv 
\, v^8 \, 
\exp\left[ \pi\Lambda \left(M - v\right)^2 
-  \pi\Lambda \, v^2 
\right]  \nonumber \\&&
\hskip 1in \left(1 - {2\over 9} \, v^2 + {64\over 99} \, v^4\right)  \,
\left(2 M v\right)^{-1}
\delta(2 v - M) \nonumber \\
&=& -{1\over 8\pi}   {\Omega_8} \, g^2\, \left({M\over 2}\right)^8\,
\left( 1 - {1\over 18} M^2 + {4\over 99} M^4\right)=
 - {47\over 264 \, \pi}\, \Omega_8\, g^2\, ,
\een
where in the last step we have used $M=2$. In \S\ref{sunitarity} we shall
argue that this result is consistent with unitarity.

We shall now show that although
each of the quantities $J_1$, $I_1$ and $I_2$ depends on the 
arbitrary parameter $\Lambda$, their sum does not
depend on $\Lambda$. 
For this, note that from \refb{edefJ1} we get
\ben \label{eL1}
{d\over d\Lambda}J_1
&=& 
-{1\over 16 \, \pi} \, M^2 \, g^2
\int d^2 \tau \int d^2 z\, 
2\,  (2\pi)^{-4} \, \left( 32 \pi^4 -32 {\pi^3 \over \tau_2} +
512 \, {\pi^2\over\tau_2^2}\right)  \nonumber \\ &&
\hskip 2in \exp[4\pi z_2 - 4\pi z_2^2/\tau_2] \, \tau_2^{-5}
\, \Theta(\tau_2 - z_2 - \Lambda) \delta(z_2-\Lambda)\, ,
\nonumber \\
&=& - 2^{-7} \pi^{-5} \, M^2 \, g^2
\int_{2\Lambda}^\infty d\tau_2 \, 
\exp[4\pi \Lambda - 4\pi \Lambda^2/\tau_2] \, \tau_2^{-5}\, 
\left( 32 \pi^4 -32 {\pi^3 \over \tau_2} +
512 \, {\pi^2\over\tau_2^2}\right) \, , \nonumber \\
\een
where in the first step we have used the $z\to \tau-z$ symmetry to 
combine two terms into a single term.
On the other hand from \refb{e1aa}, \refb{ej3i1i2} we get
\be
{d\over d\Lambda} (I_1+I_2) 
= -i\, 2^8\, \pi^8 \, M^2 \,g^2 \, \int{d^{10} k \over (2\pi)^{10}} \, 
\exp[-\pi\Lambda k^2 -
\pi\Lambda (p-k)^2 ]\, (k^2)^{-1}
(1 - 2 \, k_1^2 + 64 \, k_1^2 k_2^2)  \, 
\ee
where again we have exploited the $k\to (p-k)$ symmetry to combine two terms
into a single term.
Once the pole associated with $\{(p-k)^2\}^{-1}$ has been removed, there is
no obstruction to taking the $k^0$ integration contour to lie along the imaginary
axis, and
representing $(k^2)^{-1}$ as $\int_0^\infty ds e^{-s k^2}$. Carrying out the
integration over $k^\mu$ using the rules of gaussian integration, and defining
$\tau_2= (s+2\pi\Lambda)/\pi$, we get
\be\label{eL2}
{d\over d\Lambda} (I_1+I_2) 
=2^{-2} \pi^{-1} \, M^2 \, g^2
\int_{2\Lambda}^\infty d\tau_2 \, 
\exp[4\pi \Lambda - 4\pi \Lambda^2/\tau_2] \, \tau_2^{-5}\, 
\left( 1 - {1 \over \pi\tau_2} +
{16 \over\pi^2 \tau_2^2}\right)\, .
\ee
Using \refb{eL1} and \refb{eL2} we get
\be
{d\over d\Lambda} (J_1+I_1+I_2)=0\, .
\ee

\sectiono{Generalizations and justification using string field theory} \label{sgen}

The procedure described in the previous section can be used to compute the renormalized
mass of any massive state in heterotic or type II string theory.
For general physical states, at one loop order
one has to consider the
possibility of mixing with other physical states at the same mass
level, but not with pure gauge or
unphysical states\cite{1401.7014}, or with states at different mass level.
If we denote by $\delta M^2$ the one loop
two point function of physical states -- typically a matrix with  
both real and imaginary parts -- then the one loop propagator will be proportional
to $(k^2+M^2 + \delta M^2)^{-1}$, and its poles will be at places where
$\det(k^2+M^2 + \delta M^2)$ vanishes.

The general strategy for computing the matrix $\delta M^2$ 
will be as follows. The two point function of general on-shell external states
of mass $M$ can be brought to the form
\be \label{efullint}
\delta M^2 = \int d^2\tau \, d^2 z \, F
\ee
where $F$ is some function of $z,\bar z,\tau,\bar\tau$ describing the two point
function of the corresponding vertex operators on the torus. Let us define 
$z_1,z_2, w_1, w_2$ via
\be 
z=z_1+ i z_2, \quad w = \tau - z \equiv w_1 + i w_2\, .
\ee
The potential divergence in \refb{efullint} comes from the region of large $z_2$ and
$w_2$. If we denote by $F_0$ the part of $F$ that can give divergent contribution, then
$F_0$ has the general form
\be \label{ef0}
F_0 = \tau_2^{-5} \, 
\exp[ \pi  M^2 z_2 w_2 / \tau_2]  \sum_{m,n} e^{ 2\pi i m z_1
+ 2\pi i n w_1} e^{2\pi z_2 + 2\pi w_2}\,  A_{m,n}(z_2, w_2)\, ,
\ee
where the sum over $m,n$ runs over a finite set of integers, and $A_{m,n}$ is
a function of $z_2,w_2$ that involves a finite  
sum of products of non-negative powers of $e^{-2\pi z_2}$, $e^{-2\pi w_2}$,
and polynomial of $1/\tau_2$ and $z_2/\tau_2$.  In defining $F_0$ we shall
include in $e^{2\pi z_2+2\pi w_2}A_{m,n}$ a term proportional to 
$e^{-2\pi p z_2 - 2\pi q w_2}$ if and only if either $p$ or $q$ is negative,
or $\sqrt{2p}+\sqrt{2q}<M$, since these
are the terms that can cause potential divergence in \refb{efullint} from the large
$z_2$ and large $w_2$ region.
In \refb{ef0} the $\tau_2^{-5} \exp[ \pi w_2 z_2 M^2 / \tau_2]$ 
factor comes from the 
non-holomorphically factorized part of the correlation function
of $e^{\pm i k.X}$ factors in the vertex operators -- expressed in 
$z\to \tau-z$ invariant form. The factors of 
$e^{2\pi  i m z_1
+2\pi  i n w_1}$, $e^{2\pi z_2 + 2\pi w_2}$, 
and the powers of $e^{-2\pi z_2}$, $e^{-2\pi w_2}$ hidden in the definition
of $A_{m,n}$ come from the expansion of the holomorphically factorized pieces
in the correlation function for large $z_2$ and $w_2$. Finally
the polynomials of $1/\tau_2$ and $z_2/\tau_2$ in the
expansion of $A_{m,n}$ come from the derivatives of the term proportional to
$(z_2-w_2)^2/\tau_2$ in the Green's function $\langle X^\mu(z,\bar z) X^\nu(w,
\bar w)\rangle$.
The presence of
the explicit factor of $e^{2\pi z_2}$ and $e^{2\pi w_2}$ 
is a reflection of the presence of the
tachyon in the left-moving sector before level matching.\footnote{These factors will
be absent in type II string theories.} 
However a term proportional to $e^{2\pi z_2}$ 
(resp. $e^{2\pi w_2}$) in the expression for $F_0$
appears only when accompanied by a factor of $e^{2\pi i z_1}$ (resp.
$e^{2\pi i w_1}$), {i.e.} $A_{p,q}$ will have its expansion beginning with
the power of $e^{-2\pi z_2}$ (resp. $e^{-2\pi w_2}$) except for $p=1$ 
(resp. $q=1$). 
Therefore for $\tau_2\ge 1$, the contribution from
terms proportional to $e^{2\pi z_2}$ (resp. $e^{2\pi w_2}$)
disappears after integration over $z_1$ (resp. $w_1$).
More generally, for $\tau_2\ge 1$ integration over $z_1$
and $w_1$ will make the integral \refb{ef0}
vanish unless $m=n=0$, but we shall continue to display them for
reasons that will become clear later.

Using \refb{efullint}, \refb{ef0} we can write
\be 
\delta M^2 =J_1+J_2\, ,
\ee
where
\be 
J_1= \int d^2w \, d^2 z \, \left[F
-  F_0 \, \Theta(w_2 - \Lambda) \, \Theta(z_2-\Lambda)
\right]\, ,
\ee
\ben
&& J_2 = \int d^2w\, d^2 z \, F_0(z_1, z_2, w_1, w_2)\, \Theta(w_2 - \Lambda) \, 
\Theta(z_2-\Lambda)
\nonumber \\
&=&  \int d^2w \, d^2 z \, \Theta(w_2 - \Lambda) \Theta(z_2-\Lambda)\, \tau_2^{-5}\, 
\,  e^{\pi M^2 z_2 w_2 / \tau_2} e^{2\pi z_2+2\pi w_2} 
\sum_{m,n} e^{ 2\pi i m z_1
+2\pi i n w_1} \,  A_{m,n}(z_2, w_2)
\, , \nonumber \\
\een
and $\Lambda$ is an arbitrary constant, which we shall take to be larger than 1.
$J_1$ can be shown to be finite following the strategy used in appendix
\ref{sa}. Our strategy for evaluation of  $J_2$ will be to drop all
terms with non zero $m,n$ since they vanish by integration over $z_1$ and 
$w_1$, and for the $m=n=0$ term,
expand $e^{2\pi z_2 + 2\pi w_2}A_{0,0}$ in a power series
\be \label{eAexp}
e^{2\pi z_2 + 2\pi w_2}A_{0,0}(z_2,w_2) 
= \sum_{p,q\ge 0\atop 2\sqrt p + 2\sqrt q<M}
e^{-4\pi p z_2 - 4\pi q w_2} P_{p,q}(z_2, w_2)\, , 
\ee
where $P_{p,q}$ is a polynomial in $1/(z_2+w_2)$ and $z_2/(z_2+w_2)$.
Note that once we have focussed on terms independent of $z_1$ and $w_1$,
the series expansion is in powers of $e^{2\pi i (z-\bar z)} = e^{-4\pi z_2}$ and
$e^{2\pi i (w-\bar w)}=e^{-4\pi w_2}$.
By making the substitution\footnote{The scale factor $\pi$ is fixed
as follows. The Schwinger parameters $t_1$ and $t_2$
introduced in \S\ref{ssch} appears in the
exponent multiplied by a factor of $k^2+m^2$. On the other hand $z_2$ and $w_2$
appear in the exponent multiplied by a factor of $2\pi (L_0+\bar L_0)=\pi
(k^2+m^2)$.
}
\be
\pi \, z_2 = t_1, \quad \pi\, w_2 = t_2, \quad \pi\tau_2 = t_1+t_2\, ,
\ee
and using \refb{eAexp}, 
we can now express $J_2$ as
\be 
J_2 =\pi^3 \int_{\pi\Lambda}^\infty dt_1 \int_{\pi\Lambda}^\infty dt_2 \, (t_1+t_2)^{-5}
\sum_{p,q\ge 0\atop 2\sqrt p + 2\sqrt q<M}
\exp\left[ M^2 t_1 t_2/(t_1+t_2) - 4\, p \, t_1 - 4\, q \, t_2\right] P_{p,q}\, .
\ee
This integral diverges for $t_1,t_2\to\infty$, but we can replace this by a momentum
space integral by
comparing with the results of \S\ref{ssch}. 
Once we have made the replacement, the integration over $k^0$ has to be
interpreted as a contour integral following the procedure described in
\S\ref{sdirect}, while integration over $\vec k$ can be regarded as 
ordinary integrals running along the real axes.
This gives
finite result due to
exponential suppression factor in the integrand for large space-like momenta.  

Note that this method is
applicable for all massive states, including the ones that
do not appear as intermediate states in the scattering of massless external
states, {\it e.g.} massive states in SO(32) heterotic string theory carrying 
SO(32) spinor representation. For such states the method
of \cite{9302003,9404128} based on factorization of four point function of
massless states is not directly applicable. 
Furthermore, since this method allows us to
directly compute the one loop two point function of two arbitrary physical states
at the same mass level, we do not have to make the effort of disentangling the
contributions from different intermediate states to the four point function.

There is however a possible subtlety with this procedure arising out of the
following consideration. If we compute the
one loop two point amplitude in string field
theory, then, for sufficiently large $\Lambda$, the contribution $J_2$ comes from the sum of
Feynman diagrams of the type shown in Fig.~\ref{fqft} with different states propagating
in the loop. If we represent the Siegel gauge propagator as
\be \label{esiegel}
b_0\, \bar b_0\, (L_0+\bar L_0)^{-1}\delta_{L_0,\bar L_0}
=2\pi \, b_0\, \bar b_0\,
\int_0^\infty d\xi_2 \int_0^{1} d \xi_1\,  e^{-2\pi \xi_2(L_0+\bar L_0)}
e^{2\pi i \xi_1
(L_0-\bar L_0)}\, ,
\ee
then for the two internal  propagators of Fig.~\ref{fqft} we have two complex
variables $\xi$ and $\zeta$ -- the analog of the variable $\xi_1+i\xi_2$
in \refb{esiegel}. Now if $\xi$ and $\zeta$ could be identified 
as the moduli parameters $z$ and $w$,
then replacing the right hand side of \refb{esiegel} by the left
hand side is equivalent to 
the prescription for doing the integration in the way we
have suggested -- i.e.\ first integrate over $z_1$ and $w_1$ at fixed
$z_2$ and $w_2$, and then replace the integration over $z_2$ and $w_2$
by momentum space integrals.
However the parameters $z$
and $w$ are not directly the variables $\xi$ and $\zeta$
of the string field theory -- they are given by some functions of $\xi$ and $\zeta$.
Therefore it is not {\it a priori} guaranteed that first performing the 
integration over the real parts of $z$ and $w$, and then
treating the imaginary parts of $z$ and $w$ as Schwinger parameters 
to translate the
amplitude to a momentum space integral is a valid procedure. 
The correct procedure will be to first express the
amplitude as integrals over the variables $\xi$ and $\zeta$, carry out the integrations over
the real parts of $\xi$ and $\eta$, and then
interpret the expression as coming from momentum space integrals treating
the imaginary parts of $\xi$ and $\zeta$ 
as Schwinger parameters. We shall now
argue that this does not change the result.

Since different string field theories (related by field redefinition) lead to different
plumbing fixture variables, 
instead of focussing on any particular string field theory we shall consider the effect
of a general parameter redefinition of the form
\be
z = f(\xi, \zeta), \quad w = g(\xi,\zeta)\, .
\ee
In order to get some insight into the form of the functions $f$ and $g$,
it will be useful to recall the geometric interpretation of the parameters
$\xi$ and $\zeta$.
In string field theory the Feynman diagram of Fig.~\ref{fqft} will represent the 
effect of sewing two three punctured spheres. If the first one has punctures 
$P_1$, $P_2$, $P_3$ with local coordinates $y_1$, $y_2$ and $y_3$, and the
second one has punctures $\wt P_1$, $\wt P_2$, $\wt P_3$ with local coordinates
$\wt y_1$, $\wt y_2$ and $\wt y_3$, then the sewing is done via the relations
\be
y_2 \, \wt y_2 = e^{2\pi i\xi}, \quad y_3 \, \wt y_3 = e^{2\pi i\zeta}\, .
\ee 
The external states are inserted at the punctures $P_1$ and $\wt P_1$.
Using this geometric interpretation of the parameters $\xi$ and $\zeta$ it is
easy to see that
for large $\xi_2$ and $\zeta_2$, we have $z\simeq \xi$ and $w\simeq \zeta$.
Using this and the fact that $z$, $w$, $\xi$ and $\zeta$ are
periodic variables with period 1, we see that $z-\xi$ and $w-\zeta$ will
have expansions in non-negative powers of $e^{2\pi i \xi}$ and $e^{2\pi i\zeta}$. 
Since such a redefinition of parameters can be built from successive infinitesimal
deformations, we shall now focus on infinitesimal deformations of the form
\be  \label{eredef}
z = \xi + a(\xi,\zeta), \quad w = \zeta + b(\xi,\zeta)\, ,
\ee
where $a$ and $b$ are infinitesimal functions admitting expansion in non-negative
powers of $e^{2\pi i \xi}$ and $e^{2\pi i\zeta}$. 
If we can show that for general infinitesimal $a$ and $b$, first expressing
$J_2$ in the $\xi,\zeta$ variables
and then mapping it to momentum  space representation regarding $\xi_2$
and $\zeta_2$ as Schwinger parameters, gives the same result as what we
get by directly converting the original
expression for $J_2$
to momentum space integral treating $z_2$ and $w_2$ as Schwinger parameters,
then we would have proven a similar result for finite redefinitions relating 
$z$ and $w$ to $\xi$ and $\zeta$.
This is what we shall now show.

Taking real and imaginary parts
of \refb{eredef} we write
\ben
&& z_1 = \xi_1 + a_1(\xi_1,\xi_2,\zeta_1, \zeta_2), 
\quad z_2 = \xi_2 + a_2(\xi_1,\xi_2,\zeta_1, \zeta_2), \nonumber \\
&& w_1 = \zeta_1 + b_1(\xi_1,\xi_2,\zeta_1, \zeta_2), 
\quad w_2 = \zeta_2 + b_2(\xi_1,\xi_2,\zeta_1, \zeta_2)\, ,
\een
where $a_i$ and $b_i$ are periodic functions of $\xi_1$ and $\zeta_1$ with 
period 1.
Under this change of variables, we get
\be
J_2 = \wt J_2 + \delta J_2
\ee
where
\be
\wt J_2 = \int d^2\xi \, d^2 \zeta  \, F_0(\xi_1, \xi_2, \zeta_1,\zeta_2) \, 
\Theta(\xi_2-\Lambda) \, \Theta(\zeta_2 - \Lambda) 
\ee
and
\ben \label{emani}
\delta J_2 &=& \int d^2 \xi \, d^2 \zeta \, 
\Theta(\xi_2-\Lambda)\Theta(\zeta_2 - \Lambda) 
\sum_{i=1}^2 \left[ {\p\over \p \xi_i} \left\{
a_i \, \,
F_0\right\}  + {\p \over \p \zeta_i} \left\{b_i \, \,
F_0\right\}
\right] \nonumber \\ &&
+ \int d^2 \xi \, d^2 \zeta \, 
\left[\delta(\xi_2-\Lambda)\Theta(\zeta_2 - \Lambda) \, a_2 
+
\Theta(\xi_2-\Lambda)\delta(\zeta_2 - \Lambda) \, b_2 \right]
F_0
\, .
\een
The arguments of $a_i$, $b_i$ and $F_0$ in \refb{emani} are 
$\xi_1,\xi_2,\zeta_1,\zeta_2$.
Now $J_2$ evaluated by regarding $z_2$ and $w_2$ as Schwinger
parameters is identical to $\wt J_2$ evaluated by regarding $\xi_2$ and $\zeta_2$
as Schwinger parameters. Therefore we need to show that $\delta J_2$ evaluated
by regarding  $\xi_2$ and $\zeta_2$
as Schwinger parameters vanish. 

Since the integration rules involve carrying out 
integration over $\xi_1$ and $\zeta_1$ first at fixed $\xi_2$ and $\zeta_2$ and then
integrating over $\xi_2$ and $\zeta_2$, the derivatives with respect to
$\xi_1$ and $\zeta_1$ vanish after integration
due to the periodicity of the functions $a_i$,
$b_i$ and $F_0$ in the $\xi_1$
and $\zeta_1$ variables. Since the rest of the terms admit expansion in powers
of $e^{2\pi i \xi_1}$ and $e^{2\pi i \zeta_1}$,  
only the $\xi_1$ and $\zeta_1$ independent terms can
contribute, -- the other terms will vanish after
integration over $\xi_1$ and $\zeta_1$.  
Therefore we can write
\be \label{exx0}
\delta J_2 = \int_\Lambda^\infty d\xi_2 \int_\Lambda^\infty d\zeta_2\, 
\left[ {\p \wt a(\xi_2,\zeta_2)\over \p\xi_2} + {\p \wt b(\xi_2,\zeta_2)\over \p\zeta_2}
\right] + \int_\Lambda^\infty d\zeta_2 \, \wt a(\Lambda, \zeta_2)
+ \int_\Lambda^\infty d\xi_2 \, \wt b(\xi_2, \Lambda)\, ,
\ee
where\footnote{Note  that part of the contribution comes from 
the terms carrying powers of $e^{2\pi i z_1}$ and $e^{2\pi i w_1}$ in the
original expression for $F_0(z_1, z_2, w_1, w_2)$, since such terms, after
combining with the $\xi_1$ and $\zeta_1$ dependent terms in $a_2$ and $b_2$,
can give rise to $\xi_1$ and $\zeta_1$ independent terms in $a_2 F_0$ and
$b_2 F_0$. This is the reason we had kept such terms in the expression for
$F_0$.
}
\ben \label{exx1}
\wt a(\xi_2, \zeta_2) &\equiv& \int_0^1 d\xi_1 \int_0^1 d\zeta_1 \,
a_2(\xi_1, \xi_2, 
\zeta_1, \zeta_2) \, F_0(\xi_1, \xi_2, 
\zeta_1, \zeta_2), \nonumber \\
\wt b(\xi_2, \zeta_2) &\equiv& \int_0^1 d\xi_1 \int_0^1 d\zeta_1 \, 
b_2(\xi_1, \xi_2, 
\zeta_1, \zeta_2)\, F_0(\xi_1, \xi_2, 
\zeta_1, \zeta_2)\, .
\een
{}From the form of $F_0$ given in \refb{ef0} we see that
$\wt a$ and $\wt b$ will have expansions of the form
\be \label{eab}
\pmatrix{\wt a\cr \wt b} =
(\xi_2+\zeta_2)^{-5}\exp[\pi M^2 \xi_2 \zeta_2/(\xi_2+\zeta_2)]
\sum_{m,n\ge 0} e^{-4\pi m \xi_2 - 4\pi n \zeta_2} \pmatrix{
C^a_{m,n}\cr C^b_{m.n}}\, ,
\ee
where $C^a_{m,n}$ and $C^b_{m,n}$ are
polynomials in $1/(\xi_2+\zeta_2)$ and $\xi_2/(\xi_2+\zeta_2)$.
Formally the right hand side of \refb{exx0} vanishes by integration by parts.
However we have to remember that these are divergent integrals and in
order to make sense of them we have to replace them by momentum 
space integrals following the dictionary given in \S\ref{stoy}. Therefore
we shall now replace each of the 
terms in the expression \refb{exx0}
by momentum space integrals,
and then ask if the total contribution vanishes. 

We proceed  as follows. Using the algorithm described in
\S\ref{ssch} we first express 
$\wt a(\xi_2,\zeta_2)$ and $\wt b(\xi_2,\zeta_2)$ given in 
\refb{eab} in the form
\be 
\pmatrix{\wt a (\xi_2,\zeta_2)\cr \wt b (\xi_2,\zeta_2)}  
= \sum_{m,n\ge 0} \int{d^{10}k\over (2\pi)^{10}} 
e^{-\pi \xi_2 (k^2+4m) -\pi \zeta_2 ((p-k)^2 + 4n)} \pmatrix{f_{a,m,n}(k)\cr
f_{b,m,n}(k)}\, ,
\ee
where $f_{a,m,n}(k)$ and $f_{b,m,n}(k)$ is some polynomial in $k^0$ and
$\vec k^2$, and we have $p^2=-M^2$.
In that case $\p \wt a/\p \xi_2$ will have the expression of the form
\be \label{epr1}
{\p\wt a\over \p \xi_2} = -\pi\sum_{m,n\ge 0} \int{d^{10}k\over (2\pi)^{10}} 
e^{-\pi \xi_2 (k^2+4m) -\pi \zeta_2 ((p-k)^2 + 4n)} 
(k^2+4m) f_{a,m,n}(k)\, .
\ee
Now
the replacement rule says that after substituting the expressions given
above into the integrals appearing in \refb{exx0}, we make the replacements
\be \label{err1}
\int_\Lambda^\infty d\xi_2 e^{-\pi \xi_2 (k^2+4m)} \to {1\over \pi} 
\exp\left[-\pi \Lambda (k^2 + 4m)\right]\,
(k^2+4m)^{-1} \, ,
\ee
and
\be \label{err2}
\int_\Lambda^\infty d\zeta_2 e^{-\pi \zeta_2 ((p-k)^2+4n)} \to {1\over \pi} 
\exp\left[-\pi \Lambda \{(p-k)^2 + 4n\}\right]\,
\{(p-k)^2+4n\}^{-1} \, ,
\ee
and then interpret the integration over $k^0$
as a contour integration of the kind described in \S\ref{sdirect}, and the integration
over $\vec k$ as ordinary $(D-1)$ dimensional integral along
real axis. This makes the replacement rules \refb{err1} and \refb{err2} only formal, since
the integration over $k^0$ can run over domains in which $k^2 +4m$ or
$(p-k)^2 + 4n$ may turn negative making the left hand sides diverge.
Using these rules, we get
\ben \label{efc1}
\int_\Lambda^\infty d\xi_2 \int_\Lambda^\infty d\zeta_2\, 
{\p \wt a(\xi_2,\zeta_2)\over \p\xi_2}  
&\to& -{1\over \pi} \sum_{m,n\ge 0} \int{d^{10}k\over (2\pi)^{10}}  
\exp\left[-\pi \Lambda (k^2 + 4m)-
\pi \Lambda \{(p-k)^2 + 4n\}\right]
\nonumber \\ && \hskip 1in \times 
\{(p-k)^2+4n\}^{-1}  f_{a,m,n}(k)\, . 
\een
Note that the $(k^2+4m)^{-1}$ factor of \refb{err1}
has been cancelled by the explicit 
$(k^2+4m)$ factor produced in \refb{epr1} 
by the $\p/\p\xi_2$ operation. The right hand side of this expression is
finite, while the individual terms contributing to the left hand side can be infinite for
$M> \sqrt{4m}+\sqrt{4n}$. The rules we have proposed uses the right hand side as the
definition of the left hand side.
On the 
other hand we have
\ben \label{efc2}
\int_\Lambda^\infty d\zeta_2 \, \wt a(\Lambda, \zeta_2) 
&=& {1\over \pi}\sum_{m,n\ge 0} \int{d^{10}k\over (2\pi)^{10}}  
\exp\left[-\pi \Lambda (k^2 + 4m)-
\pi \Lambda \{(p-k)^2 + 4n\}\right]\nonumber \\ && \hskip 1in \times
\{(p-k)^2+4n\}^{-1}  f_{a,m,n}(k)\, . 
\een
Note that this is an equality -- both the left and the right hand sides are finite since the
integral of an expression of the form given in \refb{eab} is finite if either
$\xi_2$ or $\zeta_2$ is fixed. 
Therefore we
can use either description to evaluate this contribution. We now see that the right hand sides of
\refb{efc1} and \refb{efc2} cancel. A similar analysis shows that the other two terms in
\refb{exx0} also cancel. 

This shows that $\delta J_2$ vanishes. Therefore $J_2$ takes the same
value  irrespective of whether we use its expression in the $w,z$ coordinate and
express it as momentum space integral by regarding $z_2$ and $w_2$ as Schwinger
parameters, or whether we take its expression in the
$\xi,\zeta$ coordinate and express it as momentum space integral by regarding
$\xi_2$ and $\zeta_2$ as Schwinger parameters. Integrating this
result to generate finite deformations, we see that the result remains the same
irrespective of whether we use the $z,w$ variables or the sewing parameters of
a string field theory to generate the momentum space representation. Besides 
justifying the use of $z,w$ variables to generate momentum space representation, this
analysis also shows that the result is independent of which string field theory we use
to generate the momentum space representation. 

The analysis has a straightforward generalization to compactified heterotic and 
type II string theories described
by general superconformal world-sheet theories.
If we consider a vacuum with $D$ non-compact space-time dimensions, then
the overall multiplicative factor of $\tau_2^{-5}$ in \refb{ef0} will be replaced by
$\tau_2^{-D/2}$. The other difference will be that the coefficients 
$A_{m,n}$ will not only have integer powers of $e^{-2\pi z_2}$, $e^{-2\pi w_2}$,
$e^{2\pi i z_1}$, $e^{2\pi i w_1}$, but 
also fractional powers of $e^{-2\pi z_2}$ and $e^{-2\pi w_2}$. 
For example 
for compactification on  a circle of radius $R$, $A_{p,q}$ will
contain factors of 
\be
\textrm{exp}\left[-\frac{\pi i \overline{z}}{2}
\left(\frac{n}{R}+mR\right)^2+\frac{\pi i z}{2}\left(\frac{n}{R}-mR
\right)^2\right]\,,
\ee
and
\be
\textrm{exp}\left[-\frac{\pi i \overline{w}}{2}
\left(\frac{n}{R}+mR\right)^2+\frac{\pi i w}{2}\left(\frac{n}{R}-mR
\right)^2\right]\,.
\ee
Here $n,m$ are integers labelling the momentum and winding numbers 
along the circle. The rest of the analysis can be carried out as before by
converting each term into momentum space integrals.

\sectiono{Unitarity} \label{sunitarity}

In this section we shall show that the result for one loop contribution
to mass$^2$ computed using our method is consistent with unitarity. The
general analysis of \cite{1604.01783} already shows that the result 
satisfies Cutkosky rules. This would prove unitarity if in the Siegel
gauge all states with $L_0=\bar L_0=0$ had been physical states.
However in general there will also be unphysical and pure gauge states.
Hence we need to show that their contribution to the cut diagram vanishes.

While for a general amplitude establishing this requires some 
effort\cite{appear},
for the one loop two point function the analysis can be carried out as
follows. 
Let us focus on states with $L_0=\bar L_0=0$
and annihilated by $b_0$ and $\bar b_0$, since these are the states that
are associated with a cut propagator 
in the Siegel gauge. We choose a basis of states
such that unphysical states -- those not annihilated by the BRST charge
$Q_B$ -- are labelled as $|\phi_s\rangle$, and physical states -- annihilated by
$Q_B$ but not pure gauge -- are labelled as $|\chi_a\rangle$.
In this basis we do not need to introduce separately the basis of
pure gauge states. -- they can be taken to be $Q_B|\phi_s\rangle$. 
Pure gauge states have non-zero inner product only with unphysical states,
while physical states can have non-zero inner product with unphysical and
physical states.
Using the fact that the BPZ inner product is non-degenerate, one can argue
that it is possible to choose a basis in which unphysical states have non-zero
inner product only with pure gauge states and physical states have non-zero inner
product only with physical states.
We denote by $|\phi_s^c\rangle$ and $|\chi_a^c\rangle$ another basis
of unphysical and physical states, also annihilated by $b_0$,
$\bar b_0$, and
satisfying
\ben \label{einn}
&& \langle \phi_s^c|c_0^- c_0^+ Q_B |\phi_r\rangle 
=\delta_{rs}, \quad \langle\phi_s^c |c_0^- c_0^+|\phi_r\rangle=0,
\quad \langle \chi_b^c |c_0^- c_0^+|\chi_a\rangle =\delta_{ab},
\nonumber \\ &&
\langle \phi_s^c |c_0^- c_0^+|\chi_a\rangle=0, \quad
\langle\chi_b^c |c_0^- c_0^+|\phi_r
\rangle=0\, ,
\een
where
\be c_0^\pm={1\over 2} (c_0 \pm \bar c_0), \quad b_0^\pm =b_0\pm \bar b_0,
\quad L_0^\pm = L_0\pm \bar L_0\, .
\ee
{}From \refb{einn} we get 
\be \label{ensnsc}
n_s +n_s^c=3\, ,
\ee
where $n_s$ and $n_s^c$ are 
the ghost numbers of $\phi_s$ and $\phi_s^c$ respectively. 

Since Siegel gauge propagator is proportional to $b_0^+ b_0^- (L_0^+)^{-1}
\delta_{L_0,\bar L_0}$, a cut propagator in the Siegel gauge will be
proportional to $b_0^+ b_0^-\delta(L_0^+) \delta_{L_0^-, 0}$.  It is easy to see that in the
$L_0^\pm =0$ subspace,
$b_0^+ b_0^-$ may be decomposed as
\be \label{ebbd}
b_0^+ b_0^- =  |\phi_r\rangle \langle\phi_r^c|Q_B + Q_B |\phi_r\rangle \langle\phi_r^c|
+ |\chi_a\rangle \langle\chi_a^c|\, .
\ee
Now consider the diagram of Fig.~\ref{fqft} but interpret this as a string theory diagram
with all string states propagating in the internal lines. A cut passing through both
internal propagators will insert a factor of \refb{ebbd} for each propagator. Let us denote
the first one by \refb{ebbd} and the second one by
\be \label{ebbda}
 |\phi_s\rangle \langle\phi_s^c|Q_B + Q_B |\phi_s\rangle \langle\phi_s^c|
+ |\chi_b\rangle \langle\chi_b^c|\, .
\ee
We shall assume that the ket is inserted on the vertex to the left and the bra is
inserted on the vertex to the right. Now since the external state in each vertex
is physical, and since the three point function on the sphere
of two physical states and one pure
gauge state vanishes, it is easy to see that many of the contributions vanish.
For example, for the combination
\be
Q_B |\phi_s\rangle \langle\phi_s^c| \, \otimes \, |\chi_a\rangle \langle\chi_a^c| 
\ee
the left vertex will represent the three point function on the sphere of
$Q_B \phi_s$, $\chi_a$ and the external state. Since $\chi_a$ and the external
state are BRST invariant, this amplitude vanishes by standard argument
involving deformation of the BRST contour.
Only the following combination survives
from the tensor product of \refb{ebbd} and \refb{ebbda} inserted at the vertices:
\be \label{efull}
 |\chi_a\rangle \langle\chi_a^c| \otimes |\chi_b\rangle \langle\chi_b^c|
 + |\phi_r\rangle \langle\phi_r^c|Q_B \otimes Q_B |\phi_s\rangle \langle\phi_s^c|
 + Q_B |\phi_r\rangle \langle\phi_r^c| \otimes  |\phi_s\rangle \langle\phi_s^c|Q_B\, .
 \ee
Of these the first term gives the desired contribution -- the sum over physical
states.
Therefore we need to show that the contribution from the other two terms cancel.
Consider the second term. For this the left vertex has the insertion of $|\phi_r\rangle$,
$Q_B|\phi_s\rangle$ and the BRST invariant external state. We can now use the
usual argument involving deformation of the BRST contour to put $Q_B$ on the
$|\phi_r\rangle$ at the cost of getting an extra minus sign and whatever other sign
we get for passing $Q_B$ through the grassmann odd operators. Similarly for the last
term in \refb{efull}, 
the right vertex has the insertion of $\langle \phi_r^c|$, $\langle\phi_s^c|Q_B$ and
the external state, and we move $Q_B$ from $\phi_s^c$ to $\phi_r^c$. This brings
\refb{efull} to
\be \label{eman1}
 |\chi_a\rangle \langle\chi_a^c| \otimes |\chi_b\rangle \langle\chi_b^c|
 -Q_B |\phi_r\rangle \langle\phi_r^c|Q_B \otimes |\phi_s\rangle \langle\phi_s^c|
 + Q_B |\phi_r\rangle \langle\phi_r^c|Q_B \otimes  |\phi_s\rangle \langle\phi_s^c|\, .
\ee
The minus sign in the second term is due to the reversal of the orientation of the BRST
contour. No further minus signs appear since here $Q_B$ has to pass through 
$|\phi_r\rangle \langle\phi_r^c|Q_B$ which is grassmann even due to \refb{ensnsc}.
On the other hand in going from the last term in \refb{efull} to the 
the last term in \refb{eman1}, $Q_B$ has to pass through the grassmann odd
combination $|\phi_s\rangle \langle\phi_s^c|$ that gives an extra minus sign and cancels
the minus sign coming from the reversal of orientation of the BRST contour. We now see 
that the last two terms in \refb{eman1} cancel, leaving behind
the contribution from only the physical intermediate 
states in the Cutkosky rules. 
This proves unitarity of the one loop two point function.

Note that the cancelation described above involves loops carrying states of
different ghost numbers -- the ghost numbers of the states $Q_B\phi_s$ and
$\phi_s$ in \refb{efull}
differ by 1. This is a generalization of the results in ordinary 
gauge theories where the proof of unitarity in the Feynman gauge involves 
cancelation between unphysical states in the matter sector and the ghost states
propagating in the loop.

\bigskip

\noindent {\bf Acknowledgement:}
I wish to thank Bo Sundborg for raising the issue that led to this investigation,
useful discussions
and critical comments on an earlier version of the manuscript. I also thank
Roji Pius, D.~Surya Ramana and Barton Zwiebach
for useful discussions.
I thank the Pauli Center for Theoretical Studies at ETH,
Zurich, SAIFR-ICTP, Sao Paulo, Theoretical Physics group of the University
of Torino and LPTHE, Paris for hospitality
during my visit when part of this work was done.
This work was
supported in part by the 
DAE project 12-R\&D-HRI-5.02-0303 and J. C. Bose fellowship of 
the Department of Science and Technology, India.

\appendix

\sectiono{Equivalence to the $i\eps$ prescription} \label{sb}

\begin{figure}

\begin{center}

\figeps


\caption{Choice of integration contour in the complex $k^0$ 
plane that can be used to 
prove equivalence between our
prescription and the $i\eps$ prescription. \label{feps}}

\end{center}

\end{figure}

In \S\ref{sdirect} we described a specific choice of contour that  can be used to evaluate
\refb{e1}. An alternative prescription, known as the $i\eps$ prescription, is to take the
expression \refb{efinint} and define the integration over $t_1$, $t_2$
by taking the upper limits of integration to be
$t_0+i\infty$ instead of $\infty$ where $t_0$ is some fixed positive 
number\cite{berera2,1307.5124}. 
The question that
we would like to address in this appendix is: Are these two prescriptions equivalent?

As pointed out in \S\ref{ssch}, the failure of the Schwinger parameter representation is in
the use of \refb{esp1}, i.e.\ it is not possible to choose the $k^0$ contour shown in Fig.~\ref{f2} such
that $k^2+m_1^2$ and $(p-k)^2+m_2^2$ always have positive real parts so that the
integrals \refb{esp1} converge. With the new prescription of turning the contours of $t_1$ and
$t_2$ (equivalently of $s_1$ and $s_2$ in \refb{esp1}) towards 
$t_0+i\infty$, the relevant question
becomes: is it possible to deform the $k^0$ integration contours in Fig.~\ref{f2}
to a form such that
$k^2+m_1^2$ and $(p-k)^2+m_2^2$ always have negative imaginary parts? If this is the case
then the integrals in \refb{esp1} -- with the new upper limits $t_0+i\infty$ --
converge and the use of these equations will be justified.

Now since the imaginary parts of $k^2+m_1^2$ and $(p-k)^2+m_2^2$ come respectively from the
$-(k^0)^2$ and the $-(p^0-k^0)^2$ terms, in order to satisfy the requirement described above we need
$k^0$ and $p^0-k^0$ to lie either in the first quadrant or in the third quadrant. At the same time we must
ensure that the poles $Q_1$ and $Q_3$ lie to the right of the contour and the poles $Q_2$ and $Q_4$
lie to the left of the contour as in Fig.~\ref{f2}. It is easy to see that the contour shown in
Fig.~\ref{feps} satisfies these requirements. In drawing this we have used that we need to take the limit
of $p^0$ approaching the real axis from the first quadrant, and have consequently taken $p^0$ to
have a small positive imaginary part.

Note that unlike the contours shown in Fig.~\ref{f2}, the contour shown in 
Fig.~\ref{feps} does not approach $\pm i\infty$ at the two ends. Instead it
approaches $\pm i\infty$ plus finite real parts. It is easy to
see however that the integrand in \refb{e1} decays exponentially as $k^0\to A \pm i\infty$ for any
finite real $A$ and hence the contour shown in Fig.~\ref{feps} can be deformed to the ones in
Fig.~\ref{f2} without changing the value of the integral \refb{e1}.

This shows that at least for the one loop two point function the prescription 
of \cite{berera2,1307.5124}
agrees with the prescription of \cite{1604.01783} that we have used in
\S\ref{sdirect}. Whether the two prescriptions agree for
general amplitudes is not known to us at present.

\sectiono{Finiteness of $J_1$} \label{sa}

In this appendix we shall show that $J_1$ defined in \refb{edefJ1} receives a finite
contribution from the $z_2\ge \Lambda$, $w_2\equiv \tau_2-z_2\ge \Lambda$ 
region of integration. 
For this let us introduce variables
\be 
u = e^{2\pi i z}, \quad v = e^{2\pi i w} = e^{2\pi i (\tau - z)}\, .
\ee
In that case $F$ given in \refb{edelM} has the form
\be \label{eapp1}
F(z,\bar z, \tau, \bar\tau)
= \exp[4\pi w_2 z_2 /(z_2+w_2)] \, \tau_2^{-5} \, G(u,v)
\, ,
\ee
where
\ben \label{eGdef}
G(u,v) &\equiv& 
\left\{\sum_{\nu} \overline{\vt_{\nu}(0)^{16}}\right\}
(\overline{\eta(\tau)})^{-18} (\eta(\tau))^{-6} (\overline{\vt_1'(0)})^{-4}
e^{2\pi i (z-\bar z)} \left( {\vt_{1}(z)\overline{\vt_{1}(z)}}\right)^2 \nonumber \\ &&
\hskip 1in \times
\left[\left( {\overline{\vt_1'(z)}\over \overline{\vt_1(z)}}\right)^2 
- {\overline{\vt_1''(z)}\over \overline{\vt_1(z)}} - {\pi\over \tau_2}\right]^2
\een
can be organized as
\be \label{eapp2}
G(u,v) = h(u,v)  
\sum_{i=0}^2 \tau_2^{-i} \left[ a^{(i)} \bar u^{-1} \bar v^{-1}
+ \bar u^{-1} f^{(i)}(\bar v) + \bar v^{-1}
f^{(i)}(\bar u) + g^{(i)}(\bar u, \bar v)\right]\, .
\ee
Here $\tau_2$ has to be interpreted as 
$z_2+w_2$, $a^{(i)}$'s are constants, 
and $h$, 
$f^{(i)}$ and
$g^{(i)}$'s are holomorphic functions of their arguments in the domain 
$|u|<1$, $|v|<1$. The form of these functions can be easily read out from 
\refb{eGdef} and known expansions of the theta and eta functions,
{\it e.g.} we have
\ben 
&& h(u,v)= e^{2\pi i z} \vt_1(z)^2 / \eta(\tau)^6\, , \nonumber \\
&& a^{(0)} = 0, \quad a^{(1)} = 0, \quad a^{(2)} = - 2^{-3}\pi^{-2}\, , \nonumber \\
&& f^{(0)}(\bar u) = - 2 \, \bar u\, (1-\bar u)^{-2}, \quad f^{(1)}(\bar u) =
- \pi^{-1}, \quad f^{(2)}(\bar u) = 2^{-3}\pi^{-2} \, (2-\bar u)\, , \nonumber \\
&&  g^{(0)}(\bar u,\bar v) = \left\{\sum_{\nu} \overline{\vt_{\nu}(0)^{16}}\right\}
(\overline{\eta(\tau)})^{-18} (\overline{\vt_1'(0)})^{-4}
e^{-2\pi i \bar z} \left(\overline{\vt_{1}(z)}\right)^2 
\left[\left( {\overline{\vt_1'(z)}\over \overline{\vt_1(z)}}\right)^2 
- {\overline{\vt_1''(z)}\over \overline{\vt_1(z)}}\right]^2\nonumber \\
&& \hskip 2in - \bar v^{-1} f^{(0)}(\bar u) - \bar u^{-1} f^{(0)}(\bar v)
\, , \nonumber \\
&&  g^{(1)}(\bar u,\bar v) = -2\pi \left\{\sum_{\nu} \overline{\vt_{\nu}(0)^{16}}\right\}
(\overline{\eta(\tau)})^{-18} (\overline{\vt_1'(0)})^{-4}
e^{-2\pi i \bar z} \left(\overline{\vt_{1}(z)}\right)^2 
\left[\left( {\overline{\vt_1'(z)}\over \overline{\vt_1(z)}}\right) 
- {\overline{\vt_1''(z)}\over \overline{\vt_1(z)}}\right]\nonumber \\
&& \hskip 2in - \bar v^{-1} f^{(1)}(\bar u) - \bar u^{-1} f^{(1)}(\bar v)
\, , \nonumber \\
&&  g^{(2)}(\bar u,\bar v) = \pi^2 \left\{\sum_{\nu} \overline{\vt_{\nu}(0)^{16}}\right\}
(\overline{\eta(\tau)})^{-18} (\overline{\vt_1'(0)})^{-4}
e^{-2\pi i \bar z} \left(\overline{\vt_{1}(z)}\right)^2 \nonumber \\
&& \hskip 2in 
-\bar u^{-1} \bar v^{-1} a^{(2)}  - \bar v^{-1} f^{(2)}(\bar u) - \bar u^{-1} f^{(2)}(\bar v)\, .
\een
Being holomorphic inside the disks $|u|<1$, $|v|<1$,
these functions have Taylor series expansions
of the form
\be \label{etaylor}
f^{(i)}(\bar u) = \sum_{m=0}^\infty f^{(i)}_m \bar u^m, \quad 
g^{(i)}(\bar u, \bar v) = \sum_{m,n=0}^\infty g^{(i)}_{m,n} \bar u^m\bar v^n, 
\quad h(u,v) = \sum_{m,n=0}^\infty h_{m,n} u^m v^n\, .
\ee
Now after integration over $z_1$ and $w_1$, only those terms in the 
expression for $F$ which carry equal powers of $u$ and $\bar u$, and also
equal powers of $v$ and $\bar v$ will survive, This gives, using \refb{eapp1},
\refb{eapp2} and \refb{etaylor}:
\be  \label{eintzz}
\int dz_1 \int dw_1 F = \sum_{m,n\ge 0} 
\exp[4\pi w_2 z_2 /(z_2+w_2) - 4\pi m z_2 - 4\pi n w_2] \,  \tau_2^{-5}\,
\sum_{i=0}^2 A^{(i)}_{m,n} \tau_2^{-i} \, ,
\ee
where 
\be \label{eahg}
A^{(i)}_{m,n} = h_{m,n} g^{(i)}_{m,n}\, .
\ee
The $(m,n)=(0,0)$ term in \refb{eintzz} 
is subtracted away from $F$ in \refb{edefJ1}. 
It can be easily seen that the term in the argument of the
exponential in \refb{eintzz}
is always negative or 0 
for $(m,n)\ne (0,0)$ and hence for each $(m,n)\ne (0,0)$ 
the integral converges due to the
$\tau_2^{-5}$ factor. It will however be instructive to investigate the individual
terms in some more detail. The $m=1$, $n=0$ term has an  exponential
factor $\exp[-4\pi z_2^2 / (z_2+w_2)]$. If we first carry out the $z_2$ integral at fixed
$w_2$, the leading contribution for large $w_2$ comes from the
$z_2\sim w_2^{1/2}$ region, and, after carrying out the integration over $z_2$ the
integrand falls off as $w_2^{-9/2}$ for large $w_2$. The subsequent  
integral over $w_2$ gives a finite result.\footnote{If 
we had considered compactified string theory with $D$ non-compact space-time dimensions
then the $\tau_2^{-5}$ factor will be replaced by $\tau_2^{-D/2}$ and after integration
over $z_2$ is performed, the integrand will fall off as $w_2^{-(D-1)/2}$. This integral
diverges for $D\le 3$. This is related to an infrared divergence
of the diagram of Fig.~\ref{fqft} for $m_1=0$ and $m_2=M$ in $D\le 3$.}  A similar
contribution will arise from the $m=0$, $n=1$ term in \refb{eintzz}. 
For terms with $m\ge 2$, $n=0$ the $z_2$ integral yields a result of order $w_2^{-5}$ 
for large $w_2$, and the result is finite
after integration over $w_2$. Similar remark holds for the $m=0$, $n\ge 2$ 
term. Finally for terms
with $m,n\ge 1$, the integrand falls off exponentially for large $z_2$
and/or $w_2$ and the integral receives a finite contribution.

This shows that each term
in the sum in \refb{eintzz} gives finite result after integration over $z_2$ and $w_2$, but
one could still wonder if the sum over $m,n$ could lead to divergence. 
For exploring this possibility we need to know the growth rate of $A^{(i)}_{m,n}$
for large $m$ and/or $n$.  For this recall that since $h$, $f^{(i)}$ and $g^{(i)}$ are
holomorphic function of their arguments for $|u|<1$, $|v|<1$, the Taylor series expansions
\refb{etaylor} should converge in this domain.
This means that for any positive constant
$\Lambda_0$, we can find another positive constant $K$ such that
\be \label{eboa}
|f^{(i)}_m| < K e^{2\pi \Lambda_0 m}, \quad 
|g^{(i)}_{m,n}| < K e^{2\pi \Lambda_0 (m+n)}, \quad 
|h_{m,n}| < K e^{2\pi \Lambda_0 (m+n)}\, .
\ee
\refb{eahg} now gives
\be
|A^{(i)}_{m,n}| < K^2 \, e^{4\pi \Lambda_0 (m+n)}\, .
\ee
We shall take $\Lambda_0<\Lambda$.
Using this we can put the following upper bound to the  integral of the
series expansion \refb{eintzz} without the $m=n=0$ term:
\ben \label{ebound}
&& 
\int_\Lambda^\infty dz_2 \int_\Lambda^\infty dw_2 \, \sum_{m,n\ge 0\atop (m,n)\ne (0,0)} 
\exp[4\pi w_2 z_2 /(z_2+w_2) - 4\pi m z_2 - 4\pi n w_2]
\, \sum_{i=0}^2 A^{(i)}_{m,n} \tau_2^{-5-i}
\nonumber \\
&\le&
\Delta_0\, \sum_{m,n\ge 0\atop (m,n)\ne (0,0)}
\int_\Lambda^\infty dz_2 \int_\Lambda^\infty dw_2 \, 
\exp[4\pi w_2 z_2 /(z_2+w_2) - 4\pi m z_2 - 4\pi n w_2] 
\, e^{4\pi  \Lambda_0 (m+n)} \tau_2^{-5}\, , \nonumber \\
&& \hskip 1in \Delta_0\equiv K^2 \left(1+{1\over 2}\Lambda^{-1}+{1\over 4}
\Lambda^{-2}
\right)\, .
\een
We now consider the following terms separately, leaving aside the $(m,n)=(0,1)$
and $(1,0)$ terms since their contribution has been analyzed separately anyway
and found to be finite.
\begin{enumerate}
\item First consider the sum of all the terms with $m\ge 2$, $n=0$. 
In this case the sum is bounded by
\ben
&& \Delta_0 \sum_{m=2}^\infty \int_\Lambda^\infty dz_2 \int_\Lambda^\infty dw_2\, 
\exp[4\pi w_2 z_2 /(z_2+w_2) - 4\pi m z_2] 
\, e^{4\pi  \Lambda_0 m} \, \tau_2^{-5}\nonumber \\
&\le & \Delta_0 \sum_{m=2}^\infty \int_\Lambda^\infty dz_2 \int_\Lambda^\infty dw_2
\, \exp[- 4\pi (m-1) z_2 ] 
\, e^{4\pi  \Lambda_0 m} w_2^{-5}\nonumber \\
&=& {1\over 16\pi} \, \Delta_0 \, e^{4\pi \Lambda_0} \, \Lambda^{-4}\,
\sum_{m=2}^\infty {1\over m-1} e^{- 4\pi (m-1) (\Lambda-\Lambda_0)}\, .
\een
Since $\Lambda>\Lambda_0$, this is a convergent sum. This shows that in the
expression for $J_1$, the sum over $m$ for $n=0$ is convergent.
\item Sum over all terms with $m=0$, $n\ge 2$ can be dealt with similarly.
\item Finally the sum over all terms in   \refb{ebound}  with $m\ge 1$, 
$n\ge 1$, can be
bounded as
\ben \label{eleft}
&& \Delta_0 \sum_{m,n\ge 1}
\int_\Lambda^\infty dz_2 \int_\Lambda^\infty dw_2 \, 
\exp[4\pi w_2 z_2 /(z_2+w_2) - 4\pi m z_2 - 4\pi n w_2] 
\, e^{4\pi  \Lambda_0 (m+n)} \tau_2^{-5} \nonumber \\
&\le & \Delta_0 \sum_{m,n\ge 1}
\int_\Lambda^\infty dz_2 \int_\Lambda^\infty dw_2 \, 
\exp[- 4\pi (m-1) z_2 - 4\pi n w_2] \, e^{4\pi  \Lambda_0 (m+n)} z_2^{-5}
\nonumber \\
&\le & {1\over 4\pi} \, \Delta_0 
\int_\Lambda^\infty dz_2 \left[ z_2^{-5} e^{4\pi \Lambda_0}+ \sum_{m\ge 2} 
e^{- 4\pi (m-1) z_2} e^{4\pi \Lambda_0 m}
\Lambda^{-5}\right]
\sum_{n=1}^\infty {1\over n} e^{-4\pi \Lambda n} e^{4\pi  \Lambda_0n}
\nonumber \\
&=&  {1\over 4\pi} \, \Delta_0 e^{4\pi \Lambda_0} \left[
{1\over 4 \Lambda^4} +{1\over 4\pi \Lambda^5} \sum_{m\ge 2} {1\over m-1}
e^{-4\pi (\Lambda-\Lambda_0) (m-1)}\right] 
\sum_{n=1}^\infty {1\over n} e^{-4\pi (\Lambda-\Lambda_0) n} \, .
\een
Since $\Lambda>\Lambda_0$, both sums on the right hand side of 
\refb{eleft} are convergent.  This shows that the sum on the
left hand side of \refb{eleft} is also convergent.
\end{enumerate}
Combining all the results we conclude that the sum on the right hand side of
\refb{ebound} converges and hence the sum on the left hand side of 
\refb{ebound} also
converges. This in turn shows that the contribution to $J_1$ has no divergence 
from the sum over infinite set of terms.

\end{document}